\documentclass{aifrontiers}
\usepackage{aifrontiers}
\usepackage{booktabs}
\usepackage{array}
\usepackage{colortbl}
\usepackage{caption}
\usepackage{subcaption}
\usepackage{multirow}
\usepackage{multicol}
\usepackage{float}
\usepackage{enumitem}
\usepackage{xspace}
\usepackage{soul}
\usepackage{listings}
\usepackage[export]{adjustbox}
\usepackage{algorithm}
\usepackage{algorithmic}
\usepackage{latexsym}
\usepackage{placeins}
\usepackage{etoc}
\usepackage{tikz}
\usetikzlibrary{shapes,arrows,calc}
\usepackage{pifont}
\usepackage{wrapfig}
\usepackage[T1]{fontenc}
\usepackage[utf8]{inputenc}

\newlength\savewidth

\newcolumntype{x}[1]{>{\centering\arraybackslash}p{#1pt}}
\newcolumntype{y}[1]{>{\raggedright\arraybackslash}p{#1pt}}

\definecolor{baselinecolor}{gray}{0.93}

\usepackage{tcolorbox}
\definecolor{promptInk}{HTML}{263238}
\definecolor{promptRole}{HTML}{98A4C0}
\definecolor{promptRequest}{HTML}{E1997D}
\definecolor{promptExpression}{HTML}{D79ABF}
\definecolor{promptDialogue}{HTML}{76B29F}
\definecolor{promptChecklist}{HTML}{B3B3B3}

\newtcolorbox{personapromptcard}[4][]{
  colback=#2!4!white,
  colbacktitle=#2!24!white,
  colframe=#2!70!white,
  coltitle=promptInk,
  coltext=promptInk,
  boxrule=0.6pt,
  arc=1.2mm,
  left=8pt, right=8pt, top=5pt, bottom=5pt,
  toptitle=4pt, bottomtitle=4pt,
  before skip=7pt, after skip=0pt,
  fonttitle=\bfseries\fontsize{10.5}{12}\selectfont,
  fontupper=\fontsize{10}{11.8}\selectfont,
  title={\raisebox{-0.25em}{#3}\hspace{0.55em}#4},
  before upper={\raggedright\setlength{\parindent}{0pt}\setlength{\parskip}{2pt}},
  #1
}

\definecolor{infoblue}{HTML}{2A78D6}
\definecolor{actorange}{HTML}{EB6834}
\definecolor{iconink}{HTML}{3A3A38}

\newcommand{\vendorbox}[1]{\makebox[2.9ex][c]{\raisebox{-0.35ex}{#1}}\hspace{0.35em}}
\newcommand{\vendoricon}[1]{\csname vendoricon@#1\endcsname}
\expandafter\def\csname vendoricon@minimax\endcsname{\vendorbox{\includegraphics[width=2.9ex,height=2.1ex,keepaspectratio]{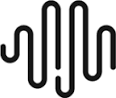}}}
\expandafter\def\csname vendoricon@kimi\endcsname{\vendorbox{\includegraphics[width=2.9ex,height=2.1ex,keepaspectratio]{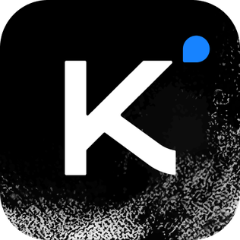}}}
\expandafter\def\csname vendoricon@muse\endcsname{\vendorbox{\includegraphics[width=2.9ex,height=2.1ex,keepaspectratio]{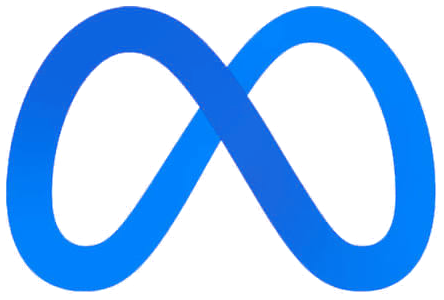}}}
\expandafter\def\csname vendoricon@grok\endcsname{\vendorbox{\includegraphics[width=2.9ex,height=2.1ex,keepaspectratio]{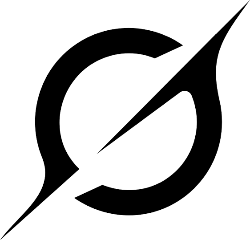}}}
\expandafter\def\csname vendoricon@qwen\endcsname{\vendorbox{\includegraphics[width=2.9ex,height=2.1ex,keepaspectratio]{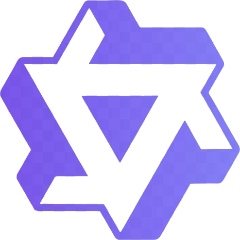}}}
\expandafter\def\csname vendoricon@gemini\endcsname{\vendorbox{\includegraphics[width=2.9ex,height=2.1ex,keepaspectratio]{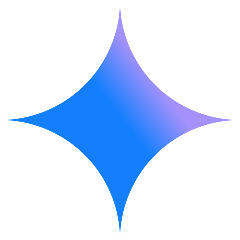}}}
\expandafter\def\csname vendoricon@gpt\endcsname{\vendorbox{\includegraphics[width=2.9ex,height=2.1ex,keepaspectratio]{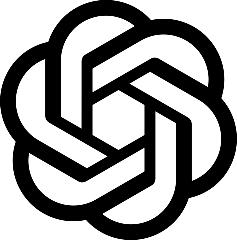}}}
\expandafter\def\csname vendoricon@deepseek\endcsname{\vendorbox{\includegraphics[width=2.9ex,height=2.1ex,keepaspectratio]{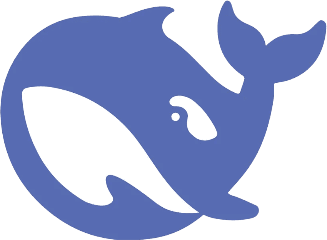}}}
\expandafter\def\csname vendoricon@claude\endcsname{\vendorbox{\includegraphics[width=2.9ex,height=2.1ex,keepaspectratio]{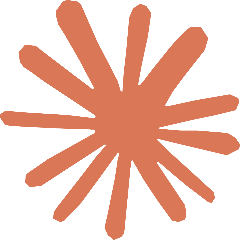}}}
\newcommand{\iconbust}{\fill (0,0.76) circle (0.21); \fill (-0.38,0) .. controls (-0.38,0.56) and (0.38,0.56) .. (0.38,0) -- cycle;}
\newcommand{\iconframe}[1]{\tikz[x=2.25ex,y=2.25ex,baseline=0.2ex,iconink,line cap=round,line join=round]{\useasboundingbox (-0.42,0) rectangle (1.0,1.1); #1}}
\newcommand{\iconexpert}{\iconframe{\iconbust \fill (-0.4,1.0) -- (0,1.14) -- (0.4,1.0) -- (0,0.86) -- cycle; \draw[line width=0.4pt] (0.33,1.0) -- (0.33,0.72);}}
\newcommand{\iconsenior}{\iconframe{\iconbust \draw[line width=0.75pt] (0.66,0.03) -- (0.66,0.62) arc (180:10:0.12);}}
\newcommand{\iconyoung}{\iconframe{\iconbust \fill[rounded corners=0.5pt] (0.54,0.06) rectangle (0.9,0.72); \fill[white] (0.6,0.2) rectangle (0.84,0.62);}}
\newcommand{\iconpractical}{\iconframe{\iconbust \draw[line width=0.55pt,fill=white] (0.7,0.36) circle (0.27); \draw[line width=0.5pt] (0.7,0.53) -- (0.7,0.36) -- (0.83,0.3);}}
\newcommand{\iconwork}{\iconframe{\fill[rounded corners=0.5pt] (-0.36,0.08) rectangle (0.62,0.72); \draw[line width=0.6pt] (-0.05,0.72) -- (-0.05,0.92) -- (0.31,0.92) -- (0.31,0.72); \draw[white,line width=0.45pt] (-0.36,0.44) -- (0.62,0.44);}}
\newcommand{\iconlife}{\iconframe{\fill (-0.42,0.56) -- (0.13,1.06) -- (0.68,0.56) -- cycle; \fill (-0.27,0.06) rectangle (0.53,0.6); \fill[white] (0.03,0.06) rectangle (0.25,0.38);}}
\newcommand{\persona}[1]{\csname icon#1\endcsname\,\textsc{#1}}
\newcommand{\cmark}{\textcolor{green!55!black}{\ding{51}}}
\newcommand{\xmark}{\textcolor{red!80!black}{\ding{55}}}
\newcommand{\tablehead}[1]{\begin{tabular}[c]{@{}c@{}}#1\end{tabular}}

\makeatletter
\newcommand{\appendixcontents}{%
  \begingroup
  \normalsize
  \setlength{\parskip}{0pt}%
  \hypersetup{linktoc=all}%
  \etocsettagdepth{main}{none}%
  \etocsettagdepth{appendix}{subsection}%
  \etocsetnexttocdepth{subsection}%
  \etocstandardlines
  \renewcommand*\l@section[2]{%
    \vspace{0.35em}%
    \@dottedtocline{1}{0em}{1.8em}{\textbf{##1}}{\textbf{##2}}}%
  \renewcommand*\l@subsection[2]{%
    \@dottedtocline{2}{1.8em}{2.6em}{##1}{##2}}%
  \etocsettocstyle{\section*{Appendix Contents}}{\par}%
  \tableofcontents
  \endgroup
}
\makeatother

\begin{document}
\etocdepthtag.toc{main}

\title{PersonaCUABench: Benchmarking Computer-Use Agents Serving Diverse Personas}

\shorttitle{PersonaCUABench}

\author{
    {\bfseries
    \mbox{Haoming Li$^{1*}$}, \mbox{Defu Cao$^{2*}$}, \mbox{Xu Hu$^{1*}$}, \mbox{Yingtao Luo$^{3}$}, \mbox{Yiyang Wang$^{4}$}, \mbox{Xianghe Xu$^{5}$}, \mbox{Jiashu He$^{6}$}, \mbox{Wanting Jiang$^{7}$}, \mbox{Yifan Zhang$^{8}$}, \mbox{Yunhan Zhao$^{9}$}, \mbox{Siyang Wu$^{10}$}, \mbox{Chen Yang$^{11}$}, \mbox{Haixin Wang$^{5}$}, \mbox{Zheyuan Deng$^{12}$}, \mbox{Taoran Li$^{13}$}, \mbox{Jingwen Zhang$^{11}$}, \mbox{Zhejian Peng$^{14}$}, \mbox{Shushu Liang$^{11}$}, \mbox{Yue Zhao$^{2}$}, \mbox{Bingzhe Li$^{1}$}, \mbox{Feng Chen$^{1}$}, \mbox{Yifan Wang$^{15}$}, \mbox{Jintao Huang$^{16}$}, \mbox{Yijun Wang$^{11}$}, \mbox{Yuexing Hao$^{17\dagger}$}, \mbox{Xiaomin Li$^{11\dagger}$}} \\[0.6em]
    {\small
    \mbox{$^{1}$The University of Texas at Dallas} \quad
    \mbox{$^{2}$University of Southern California} \quad
    \mbox{$^{3}$Carnegie Mellon University} \quad
    \mbox{$^{4}$Georgia Institute of Technology} \quad
    \mbox{$^{5}$University of California, Los Angeles} \quad
    \mbox{$^{6}$University of Pennsylvania} \quad
    \mbox{$^{7}$Amherst College} \quad
    \mbox{$^{8}$Vanderbilt University} \quad
    \mbox{$^{9}$University of California, Irvine} \quad
    \mbox{$^{10}$University of Chicago} \quad
    \mbox{$^{11}$Harvard University} \quad
    \mbox{$^{12}$Brown University} \quad
    \mbox{$^{13}$Texas A\&M University} \quad
    \mbox{$^{14}$University of Illinois Urbana-Champaign} \quad
    \mbox{$^{15}$Purdue University} \quad
    \mbox{$^{16}$The Ohio State University} \quad
    \mbox{$^{17}$Massachusetts Institute of Technology}}
}
\date{\today}
\renewcommand{\thefootnote}{\fnsymbol{footnote}}

\begin{abstract}
Computer-use agents (CUAs) are commonly evaluated against fixed task specifications, but users may omit requirements, add constraints, or require approval before consequential actions. Completing the benchmark instruction does not guarantee serving the person who asked. We introduce PersonaCUABench, which holds 74 tasks across 11 domains fixed and varies only who asks. A persona agent, playing one of 44 personas in four groups, rewrites each instruction in that user's own words and answers the CUA's questions, while the CUA sees only the dialogue, never the profile or the original instruction. Across nine backbones (3,330 runs), mean task completion falls by 36\% relative to the original requests. The losses have two sources: in 59.9\% of cells that lose points, the agent engages with what the user asked for and still gets it wrong, and 11.2\% are stops the user requested that the checker scores as failures. Even when a persona request is fully completed, the user gives the top satisfaction rating in only 62.7\% of cells, against 84.1\% for the original requests. CUAs should therefore be evaluated not only on task completion but also on whether they meet the requirements and constraints users express.

\noindent\textbf{Code:} \url{https://github.com/MatrAIx-ai/PersonaCUABench}\\
\textbf{Project:} \url{https://matraix.ai/research/persona-cua.html}
\end{abstract}

\maketitle
\begingroup
\renewcommand{\thefootnote}{}
\footnotetext{$^{*}$Co-first authors. $^{\dagger}$Correspondence: Yuexing Hao (\texttt{yuexing@mit.edu}) and Xiaomin Li (\texttt{xiaominli@g.harvard.edu}).}
\endgroup
\renewcommand{\thefootnote}{\arabic{footnote}}

\section{Introduction}

\begin{figure*}[t]
  \centering
  \includegraphics[width=\textwidth]{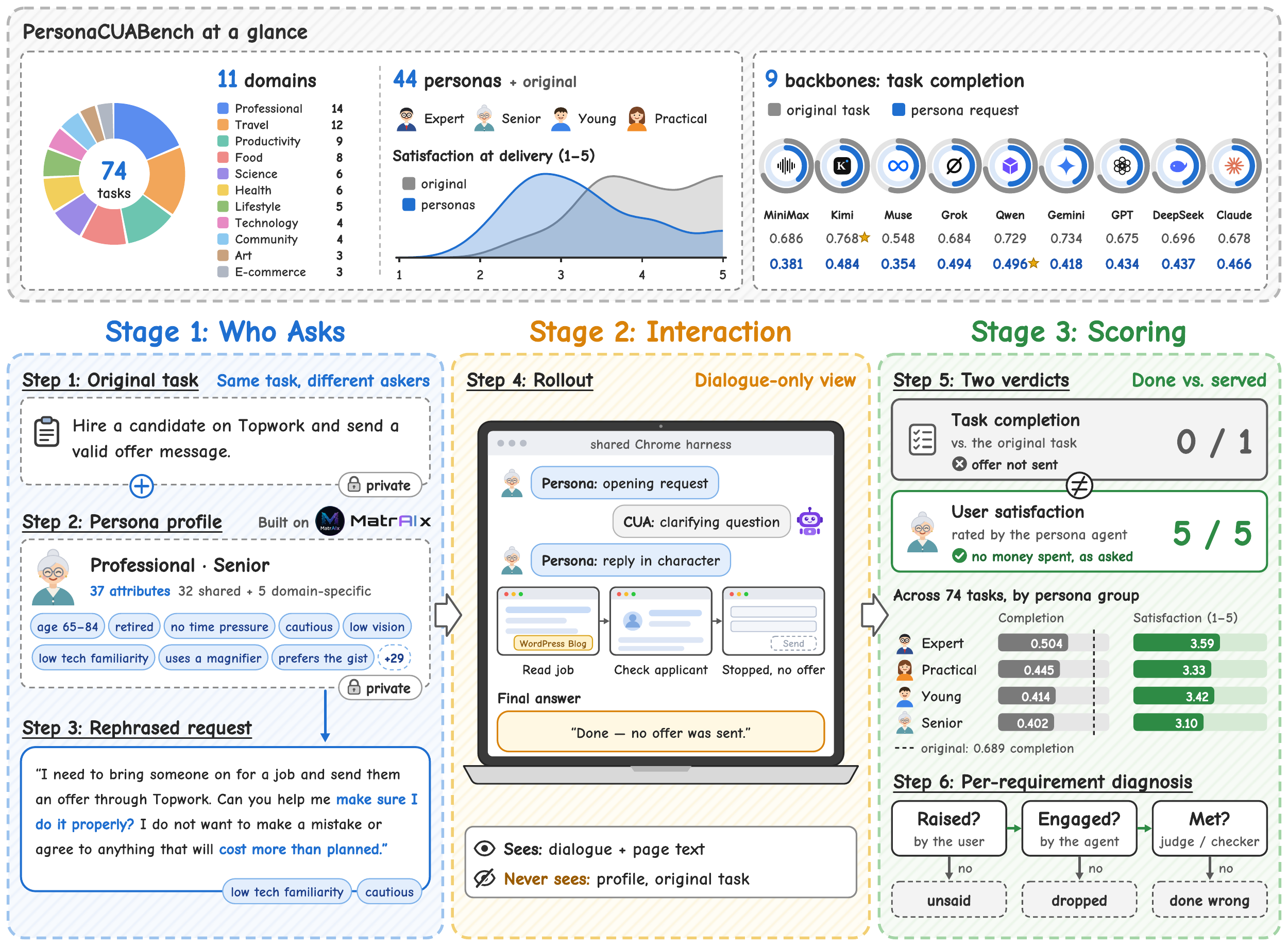}
  \caption{Overview of PersonaCUABench.
\textbf{Top:} 74 tasks in 11 domains, 44 personas, four groups in each of 11 domains,
plus the original request, task completion of nine backbones on original and persona
requests, and the distribution of user satisfaction.
\textbf{Stage 1:} a MatrAIx persona agent with a 37-attribute profile rephrases the
original task in the user's own words. The CUA sees neither the profile nor the task.
\textbf{Stage 2:} the CUA sees only the dialogue, acts in Chrome, and may ask the
persona agent questions.
\textbf{Stage 3:} each rollout gets two scores: task completion against the original
task, and satisfaction rated by the same persona agent. Here MiniMax M3 stops before
sending the offer, so the task fails while the user rates it 5/5. Each requirement
is then checked as raised, engaged, and met to locate the loss.}
  \label{fig:overview}
  \vspace{-0.2in}
\end{figure*}

Computer-use agents (CUAs) increasingly perform multi-step tasks in web browsers and desktop applications, from gathering information to making reservations and sending messages~\citep{real}. Existing benchmarks primarily evaluate whether agents complete predefined tasks~\citep{webarena, osworld}, including increasingly challenging long-horizon settings~\citep{odysseys, osworld2}. 
In real-world use, however, users may express goals incompletely, introduce personal constraints, or request confirmation before consequential actions. Satisfying a predefined task specification can therefore differ from serving the person who made the request. A fixed specification cannot reveal this gap, which raises our question: \textit{when different people ask a CUA for the same task in their own words, does it serve each of them, or only the task?}

Serving the person is harder than completing a fixed specification. The agent must infer what the user leaves unsaid and judge when to ask or confirm before acting, and for a CUA these judgments are consequential: a misread request can mean money spent, an email sent, or more disclosed than the user intended~\citep{agentcibench, hu2026trustworthycua}. Success is also user-dependent. An agent that drafts an offer and stops may fail a checker written for sending it, yet do exactly what a cautious user asked. Completing the specified task and serving the person can thus come apart, so an evaluation of CUAs needs to measure both.

Prior work makes agents personal by giving the agent stored information about the user: a logged profile, a long-term history, or a logged-in environment~\citep{personalwab, persona2web, mypcbench}. We give the agent no profile. It learns who the user is only from the request and from the user's answers to its questions. Each user is built from attributes of the MatrAIx persona schema~\citep{li2026matraixsimulatingworld83}. Holding the task fixed and changing only who asks isolates the effect of who is asking.

We introduce \textbf{PersonaCUABench} (Figure~\ref{fig:overview}), built around a persona-conditioned request protocol: a persona agent is given a profile and the original instruction, writes the opening request in that person's own words, and answers the CUA's clarifying questions in character, so the CUA works only from the dialogue. The tasks come from existing benchmarks so that outcomes stay independently checkable: information tasks on live websites and action tasks on sandboxed replicas of real sites, across everyday and work scenarios in 11 domains. Each task is asked by the four personas of its domain, one per group (\textsc{expert}, \textsc{senior}, \textsc{young}, \textsc{practical}), and by an original request that passes the instruction unchanged, without a persona. Because a request can leave requirements unsaid, we score whether the user raised each requirement, whether the agent engaged with it, and whether it was met, and we report this diagnosis alongside task completion and the user's own satisfaction.

 Evaluating nine backbones in one shared browser harness, we find that a persona-phrased request lowers average task completion from 0.689 to 0.441 with every backbone losing between 28\% and 45\%. The diagnosis locates two sources of this loss: in 59.9\% of the cells that lose points the agent {engages with} what the user {raised} yet still gets a requirement wrong, while another 11.2\% are stops the user asked for that the original checker counts as failures. We further find that task completion and user satisfaction diverge in both directions
and that{, on GPT-6 Astra,} neither revealing the user's profile nor requiring a clarification question is sufficient to recover the loss in task completion. PersonaCUABench thus provides a controlled testbed for personal CUAs: it compares different personas on the same tasks, diagnoses how user requirements are handled, and measures task completion and user satisfaction.

\section{Serving, Not Just Solving}
\label{sec:problem}

The goal of a personal computer-use agent is to serve the person who made the request. We make this precise by formalizing the three stages of Figure~\ref{fig:overview}: who asks for a task (Stage~1, \S\ref{sec:setting}), how the user and the agent interact (Stage~2, \S\ref{sec:rollout}), and how the outcome is scored (Stage~3, \S\ref{sec:eval}).

\subsection{Who Asks: Tasks, Users, and Requests}
\label{sec:setting}

Stage~1 starts from a task and decides who asks for it. Each task $t$, taken from an existing benchmark whose outcome can be checked automatically, is a triple $\bigl(x_t,\; \mathcal{R}_t,\; C_t\bigr),$
where $x_t$ is the original instruction of that benchmark, $\mathcal{R}_t= \{r_{t,1}, \dots, r_{t,n_t}\}$ its set of requirements, and $C_t$ its evaluation procedure, both defined in \S\ref{sec:eval}.

A user is a persona $p$ whose attribute values $\mathbf{v}_p$ are rendered as a natural-language profile $\pi(\mathbf{v}_p)$ (\S\ref{sec:construction}). A persona agent $U$ is given the profile and the original instruction $x_t$. The instruction describes the outcome that would fully satisfy the task, while the profile determines how the user expresses that goal. The persona agent generates the opening request
\begin{equation}
u_0 \sim U\bigl(\cdot \mid \pi(\mathbf{v}_p),\, x_t\bigr).
\label{eq:request}
\end{equation}
The original request uses the instruction directly, $u_0 = x_t$.
In persona conditions, the CUA receives only this request, while the original instruction remains private to the persona agent.%

\subsection{Interaction}
\label{sec:rollout}

Stage~2 is the episode, in which the CUA works from the dialogue alone. Starting from $h_0 = (u_0)$, the agent $M$ issues one tool call per step, $a_k \sim \pi_M(\cdot \mid h_k)$, and appends its result to the history, $h_{k+1} = h_k \oplus (a_k, o_k)$; one {agent} turn may issue several steps. Browser tools return page observations, $o_k = \mathrm{Env}(a_k)$ (Appendix~\ref{app:harness}), and \texttt{ask\_user} returns the persona agent's reply to a clarification question $q_j$,
\begin{equation}
u_j \sim U\bigl(\cdot \mid \pi(\mathbf{v}_p),\, x_t,\, \delta_{<j},\, q_j\bigr),
\label{eq:reply}
\end{equation}
which draws on $x_t$ for task details the opening request left out and on the profile for anything the instruction leaves open. The episode ends when the agent calls $\texttt{finish}(y)$ or reaches its budget of {agent} turns (\S\ref{sec:setup}), leaving the rollout record
\begin{equation}
\tau = \bigl(\delta,\; (a_k,o_k)_{k<T},\; y,\; \Omega,\; s_T\bigr)
\label{eq:rollout}
\end{equation}
of the dialogue, the tool calls and observations, the final answer, the tabs left open, and, for action tasks, the final site state.

\subsection{Scoring: Raised, Engaged, Met}
\label{sec:eval}

Stage~3 scores each rollout in two ways, task completion and user satisfaction, and labels each requirement to locate where completion is lost. Appendix~\ref{app:eval} gives the full definitions and prompts.

\textbf{Task completion.} Each requirement $r_i \in \mathcal{R}_t$ is one checkable condition on the outcome, and the task's evaluation procedure $C_t$ decides whether it was met, $m_i \in \{0,1\}$. Task completion is the share of requirements met, $\frac{1}{n_t}\sum_{i} m_i$. When $C_t$ checks the task as a whole, as the REAL~\citep{real} state checker does, every requirement shares its single verdict.

\textbf{Intent.} Completion alone does not show where a requirement was lost. For each requirement we also record whether the user raised it in the dialogue and whether the agent engaged with it, that is, acted on it, correctly or not. With met, these labels place each lost requirement at one of three points: the user never raised it, the agent never engaged with it, or the agent engaged with it and still missed it (unsaid, dropped, and done wrong).

\textbf{User satisfaction.} After the episode, the persona agent that made the request rates, in character, its satisfaction at delivery on a 1--5 scale and whether it believes the task was done. On action tasks it rates again after verification, when it is shown what changed on the site.

\section{PersonaCUABench}
\label{sec:construction}

We build the inputs of Stage~1 in Figure~\ref{fig:overview}: the task set (\S\ref{sec:tasks}) and the personas (\S\ref{sec:personas}), built independently and paired by domain, and the request each pairing produces (\S\ref{sec:rewrite}). The figure traces one pairing, a hiring task on a replica of a freelancing site asked by a \textsc{senior} user.

\subsection{Task Sources and Curation}
\label{sec:tasks}
\begin{wraptable}{R}{0.48\textwidth}
\centering
\small
\setlength{\tabcolsep}{4pt}
\caption{Task composition. Three domains (travel, food, lifestyle) contain tasks from both sources.}
\label{tab:tasks}
\begin{tabular}{@{}lccc@{}}
\toprule
 & \textbf{Odysseys} & \textbf{REAL} & \textbf{Total} \\
\midrule
Tasks & 38 & 36 & 74 \\
Task type & Information & Action & --- \\
Environment & Live web & Replicas & --- \\
Requirements & 206 & 107 & 313 \\
Completion check & Rubric judge & State checker & --- \\
Life / work tasks & 28 / 10 & 12 / 24 & 40 / 34 \\
Domains & 9 & 5 & 11 \\
\bottomrule
\end{tabular}
\end{wraptable}
We draw tasks from two recent benchmarks that together cover what users ask computer-use agents to do: find information and take actions. Odysseys~\citep{odysseys} provides long-horizon information tasks on live websites, in which the agent gathers, compares, and reports information. REAL provides action tasks on deterministic replicas of widely used websites, such as ordering food, sending an email, or booking a stay. Both remain hard for current agents, whose best results complete fewer than half of the tasks, and both check outcomes without a human: a rubric judge that never sees the dialogue scores Odysseys, and a programmatic check of the final site state scores REAL. Task completion therefore stays comparable across users who phrase the same task differently. From the two sources we keep tasks that run reproducibly in our harness and leave room for different users to want different things, dropping pure navigation, sites that block automated access, and duplicates (Appendix~\ref{app:tasks}). As shown in Table~\ref{tab:tasks}, this leaves 74 tasks, 38 information and 36 action, drawn as subsets of the 200 Odysseys and 112 REAL tasks and spread across 11 domains and two scenarios (40 life and 34 work). Figure~\ref{fig:overview} (top) shows how they spread over the domains.

\subsection{Persona Design and Assignment}
\label{sec:personas}
\begin{wraptable}{R}{0.48\textwidth}
\centering
\footnotesize
\setlength{\tabcolsep}{1.25pt}
\renewcommand{\arraystretch}{1.15}
\caption{The four persona groups $\mathcal{K}$, each instantiated in all 11 domains, with their values on the five grouping attributes. All 11 personas of a group share these values, with five exceptions listed with the attributes.}
\label{tab:roles}
\begin{tabular}{@{}
>{\centering\arraybackslash}m{44.5pt}
>{\centering\arraybackslash}m{23pt}
>{\centering\arraybackslash}m{40pt}
>{\centering\arraybackslash}m{29pt}
>{\centering\arraybackslash}m{35.2pt}
>{\centering\arraybackslash}m{\dimexpr\linewidth-184.2pt\relax}@{}}
\toprule
\textbf{Group} & \textbf{Age} & \textbf{Education} & \textbf{Income} & \textbf{Seniority} & \textbf{Time} \\
\midrule
\shortstack{\iconexpert\\\textsc{expert}} & 25--54 & Master's & \shortstack{\$100k--\\200k} & Senior+ & Moderate \\
\midrule
\shortstack{\iconsenior\\\textsc{senior}} & 65--84 & Secondary & \shortstack{\$25k--\\50k} & Retired & No rush \\
\midrule
\shortstack{\iconyoung\\\textsc{young}} & 18--34 & \shortstack{Some\\college} & $<$\$25k & \shortstack{Student/\\entry} & Moderate \\
\midrule
\shortstack{\iconpractical\\\textsc{practical}} & 25--54 & Bachelor's & \shortstack{\$50k--\\100k} & \shortstack{Mid--\\manager} & Deadline \\
\bottomrule
\end{tabular}
\end{wraptable}
To capture user differences in age, education, income, seniority, and time pressure, we define four persona groups $\mathcal{K} = \{\textsc{expert}, \textsc{senior}, \textsc{young}, \textsc{practical}\}$ (Table~\ref{tab:roles}): domain experts, older adults, young adults, and practical users who work to a deadline. We instantiate every group in each of the 11 domains $\mathcal{F}$, setting the user's background per domain (e.g., the travel \textsc{practical} persona is a recent immigrant facing a language barrier). The persona of group $\kappa \in \mathcal{K}$ in domain $f \in \mathcal{F}$ is $p_{f,\kappa}$, giving $|\mathcal{F}| \times |\mathcal{K}| = 44$ personas.

Each persona is described by 37 attributes selected from the 1,290-dimensional MatrAIx schema: 32 general attributes shared by all personas and five specific attributes chosen per domain (Figure~\ref{fig:overview}, {Step~2}, shows one \textsc{senior} profile). The general attributes cover factors known to shape how people use the web and make decisions~\citep{ghose2013mobile, schwartz2002maximizing, lastovicka1999lifestyle, weber2002domain}, organized around four questions: whether users can use the result, what they ask for and expect, which trade-offs they accept, and who they are. Domain-specific attributes capture skills, familiarity, and interests, such as travel experience or time-management habits. 

Five general attributes define the groups: age bracket, highest education, household income, seniority, and time pressure. Seniority, together with technology savviness and domain-specific skills, reflects expertise. These axes are among the strongest known predictors of differences in technology use~\citep{Czaja2006FactorsPT, hargittai2001secondleveldigitaldividemapping, vandeursen2014digital} and decision making~\citep{ordonez1997decisions}. Within a group they take nearly fixed values (Table~\ref{tab:roles}); Appendix~\ref{app:attributes} lists all 37 attributes. We render the attribute values $\mathbf{v}_p$ of persona $p$ as a structured natural-language profile $\pi(\mathbf{v}_p)$.

Each task $t$ belongs to one domain $\phi(t) \in \mathcal{F}$ and is paired with the four personas of that domain, one from each group, together with the original request $p_{\varnothing}$:
\begin{equation}
\mathcal{P}(t) = \bigl\{ p_{\phi(t),\kappa} : \kappa \in \mathcal{K} \bigr\} \cup \{ p_{\varnothing} \}.
\label{eq:pairing}
\end{equation}

\begin{table*}[t]
\centering
\fontsize{9}{11}\selectfont
\setlength{\tabcolsep}{3pt}
\caption{Task completion and user satisfaction at delivery over all 74 tasks. Models are listed in order of release date, and bold marks the best model in each column. $^{\dagger}$ denotes open-source models.}
\label{tab:main}
\renewcommand{\arraystretch}{1.12}%
\setlength{\aboverulesep}{0pt}\setlength{\belowrulesep}{0pt}%
\resizebox{\textwidth}{!}{%
\begin{tabular}{@{}l|c|c|c|c|c|c|c|c|c|c@{}}
\toprule
\multicolumn{1}{c|}{\multirow{2}{*}{Model}} & \multicolumn{5}{c|}{\rule{0pt}{2.6ex}\textbf{Task Completion}} & \multicolumn{5}{c}{\textbf{Satisfaction} (1--5)} \\
\cmidrule(lr){2-6}\cmidrule(l){7-11}
\multicolumn{1}{c|}{} & \rule{0pt}{2.6ex}Original & \textsc{expert} & \textsc{senior} & \textsc{young} & \textsc{practical} & Original & \textsc{expert} & \textsc{senior} & \textsc{young} & \textsc{practical} \\
\midrule
\vendoricon{minimax}MiniMax M3 & 0.686 & 0.447 & 0.334 & 0.355 & 0.386 & 4.01 & 3.08 & 3.05 & 3.09 & 3.09 \\
\vendoricon{kimi}Kimi K3 & \textbf{0.768} & \textbf{0.555} & 0.432 & 0.452 & 0.496 & 4.24 & \textbf{3.88} & 3.22 & 3.57 & 3.47 \\
\vendoricon{muse}Muse Glimmer 30B$^{\dagger}$ & 0.548 & 0.413 & 0.312 & 0.374 & 0.316 & 3.59 & 3.31 & 2.88 & 3.16 & 2.85 \\
\vendoricon{grok}Grok 4.6 & 0.684 & 0.543 & \textbf{0.476} & 0.443 & \textbf{0.515} & 4.16 & \textbf{3.88} & 3.14 & 3.62 & 3.53 \\
\vendoricon{qwen}Qwen 3.8 27B$^{\dagger}$ & 0.729 & 0.554 & 0.431 & \textbf{0.497} & 0.501 & 4.09 & 3.77 & \textbf{3.38} & \textbf{3.66} & 3.26 \\
\vendoricon{gemini}Gemini 3.8 Flash & 0.734 & 0.476 & 0.360 & 0.386 & 0.451 & \textbf{4.33} & 3.34 & 2.86 & 3.32 & 3.33 \\
\vendoricon{gpt}GPT-6 Astra & 0.675 & 0.504 & 0.375 & 0.399 & 0.457 & 4.01 & 3.74 & 3.03 & 3.42 & \textbf{3.55} \\
\vendoricon{deepseek}DeepSeek V4.1 Flash & 0.696 & 0.485 & 0.444 & 0.366 & 0.452 & 4.00 & 3.64 & 3.20 & 3.53 & 3.45 \\
\vendoricon{claude}Claude Opus 5.5 & 0.678 & 0.550 & 0.439 & 0.444 & 0.433 & 3.89 & 3.69 & 3.15 & 3.43 & 3.49 \\
\midrule
\textbf{Mean} & 0.689 & 0.504 & 0.402 & 0.414 & 0.445 & 4.04 & 3.59 & 3.10 & 3.42 & 3.33 \\
\bottomrule
\end{tabular}}

\end{table*}

\subsection{Request Construction}
\label{sec:rewrite}
\begin{wraptable}{R}{0.48\textwidth}
\centering
\small
\setlength{\tabcolsep}{2pt}
\caption{Task completion by scenario and persona group, mean over nine models. Table~\ref{tab:slices} gives the intervals. }
\label{tab:groups}
\begin{tabular*}{\linewidth}{@{\extracolsep{\fill}}l c c c c@{}}
\toprule
\multicolumn{1}{c}{\tablehead{\textbf{Group}}} & \tablehead{\textbf{Original}} & \tablehead{\textbf{Persona}} & \tablehead{\textbf{Absolute}\\\textbf{change}} & \tablehead{\textbf{Relative}\\\textbf{change (\%)}} \\
\midrule
All tasks (74)          & 0.689 & 0.441 & $-0.248$ & $-36$ \\
\iconwork\,Work (34)    & 0.705 & 0.431 & $-0.274$ & $-39$ \\
\iconlife\,Life (40)    & 0.675 & 0.449 & $-0.226$ & $-33$ \\
\midrule
\persona{expert}    & 0.689 & 0.504 & $-0.185$ & $-27$ \\
\persona{senior}    & 0.689 & 0.402 & $-0.287$ & $-42$ \\
\persona{young}     & 0.689 & 0.414 & $-0.275$ & $-40$ \\
\persona{practical} & 0.689 & 0.445 & $-0.243$ & $-35$ \\
\bottomrule
\end{tabular*}
\end{wraptable}
For each pairing, the persona agent rewrites the original instruction $x_t$ into the opening request $u_0$ that the agent receives, describing the user's situation and goal in language consistent with the profile $\pi(\mathbf{v}_p)$. Domain familiarity shapes whether it names a specific website or technical step, reading preferences shape its vocabulary and length, and personal circumstances shape which details it mentions, such as a budget, a schedule, or a device. As a result, $u_0$ may state a requirement outright, describe it only loosely, omit it, or add a constraint the original instruction never specified. In Figure~\ref{fig:overview}, for instance, the \textsc{senior} user adds a constraint: confirm a fixed price before anything is sent. A requirement left out of $u_0$ can still surface later, when the agent asks a clarification question (\S\ref{sec:rollout}). The details regarding the persona agent's prompt are described in Appendix~\ref{app:persona-prompt}. 

\section{Experiments}
\label{sec:setup}

\subsection{Setup}
\label{sec:protocol}
Our experiments are a controlled comparison in which only two things vary, the execution backbone and the persona condition, while the tasks, the user model, the environment, and the scoring are held fixed. Table~\ref{tab:main} lists the nine backbones that serve as the agent under test, in order of release: MiniMax M3, Kimi K3, Muse Glimmer 30B, Grok 4.6, Qwen 3.8 27B, Gemini 3.8 Flash, GPT-6 Astra, DeepSeek V4.1 Flash, and Claude Opus 5.5. Each is run on the same $370$ cells of Eq.~\ref{eq:pairing}: every task under its four domain personas and under the original request.

{Everything else is held constant.} At the judge stage, GPT-5.6-terra fills two separate roles. One instance plays every persona agent and rates its own satisfaction in the same conversation after the episode. A fresh instance, which shares no context with the persona agent, serves as the objective judges, scoring each requirement against the rubric, and labels the loss cells. 
All nine run in one harness with the same text tools and the same budget, at most $K=100$ agent turns and $Q=5$ calls to \texttt{ask\_user} per episode, with the agent and the persona agent both sampled at temperature 1.0, and we log the tokens and API cost of every call. Appendices~\ref{app:setup-eval} further describes the harness, the cost, and the human annotation that checks the judges.

\subsection{Overall Results}
\label{sec:res-drop}

\begin{wrapfigure}{R}{0.48\textwidth}
\centering
\includegraphics[width=\linewidth]{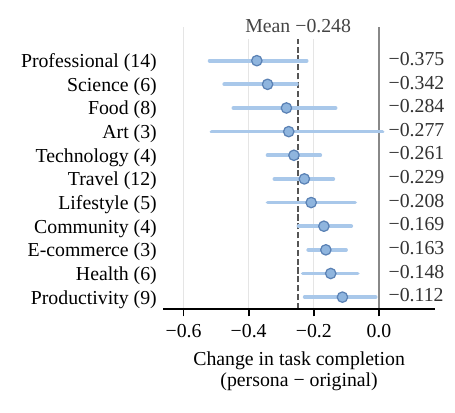}
\caption{Change in task completion by domain, mean over nine models with 95\% intervals.}
\label{fig:domains}
\end{wrapfigure}
Table~\ref{tab:main} compares the original instruction with the four persona groups. {Under every model, each persona group's task completion lies below that of the original request.} Averaged over models, task completion falls from 0.689 to 0.504 for \textsc{expert} users, 0.402 for \textsc{senior} users, 0.414 for \textsc{young} users, and 0.445 for \textsc{practical} users. The drop holds in every slice. All nine models lose in both work and life scenarios, and at least eight of nine lose in every domain, as Table~\ref{tab:groups} and Figure~\ref{fig:domains} show. Table~\ref{tab:main} also gives the user's satisfaction, the 1--5 rating that each persona agent gives its own rollout at delivery (\S\ref{sec:eval}). Averaged over models the satisfaction is 3.59, 3.10, 3.42, and 3.33 for \textsc{expert}, \textsc{senior}, \textsc{young}, and \textsc{practical} users, lowest for \textsc{senior}. The original request, which is rated without a profile, receives 4.04.

Three comparisons organize the table. \textsc{Expert} requests are the highest persona column for every model, and that column still lies 0.185 below the original request. Inside a single model, the highest and lowest persona groups differ by 0.10 to 0.13. Across models, one persona group spans 0.14 for \textsc{expert} users and 0.16 for \textsc{senior} users, so who asks moves the score on the same scale as which model runs. The ranking also changes with the group: Gemini places near the top on original requests, at 0.734, and seventh on \textsc{senior} requests, at 0.360. Each model's relative loss lies between 28\% and 45\%, and every 95\% interval in Table~\ref{tab:main-delta} lies below zero. The absolute drop grows with the original-request score ($r=-0.60$), while the relative drop does not ($r=-0.14$).

\section{Analysis}
\label{sec:results}

\begin{figure*}[tp]
\centering
\includegraphics[width=0.9\textwidth]{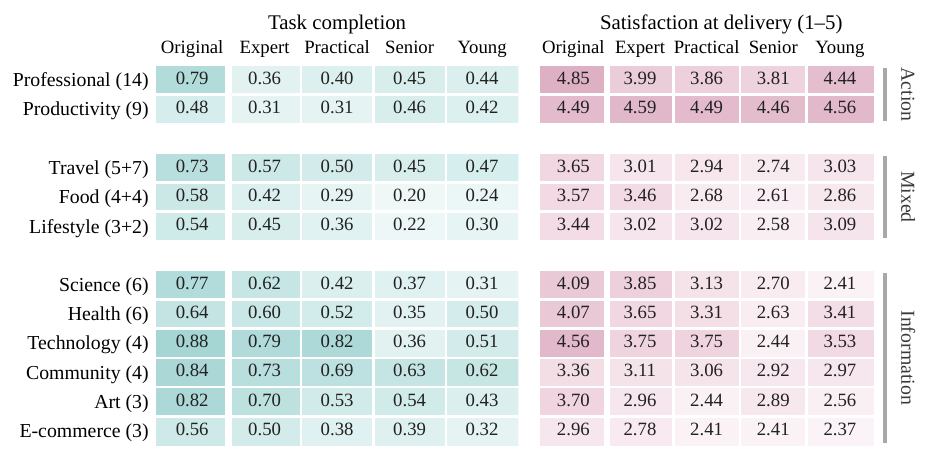}
\caption{Task completion and satisfaction at delivery by domain, for the original request and each persona group, mean over nine models. Mixed domains list their information and action tasks.}
\label{fig:domain-persona}
\vspace{-0.1in}
\end{figure*}

Section~\ref{sec:res-drop} showed that every model loses about a third of its task completion when a persona makes the request. We examine this loss through a case study and three analyses: \S\ref{sec:res-report} compares task completion with the users' own satisfaction; \S\ref{sec:res-form} traces where the drop comes from; and \S\ref{sec:res-profile} tests whether showing the user's profile or requiring the agent to ask a clarification question recovers the loss. Appendix~\ref{app:results} complements these analyses with the full breakdowns behind Table~\ref{tab:groups} and Figure~\ref{fig:domains}, and an analysis of clarification behavior showing that only 24\% of the agents' 609 questions target a requirement the user omitted.

\subsection{Task Completion vs. User Satisfaction}
\label{sec:res-report}

\begin{wrapfigure}{R}{0.48\textwidth}
\centering
\includegraphics[width=\linewidth]{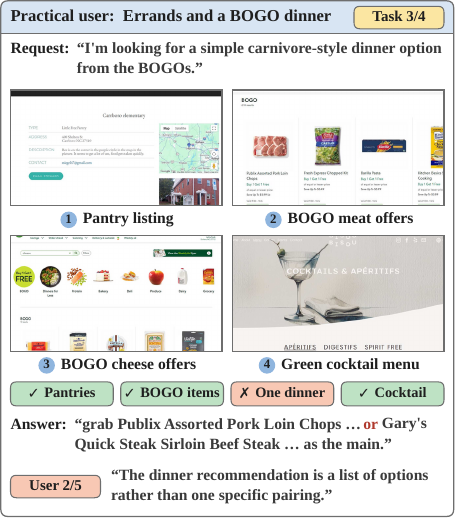}
\caption{A \textsc{practical} user's case, run by DeepSeek V4.1 Flash. Appendix~\ref{app:cases} gives one case for each of the other three groups.}
\label{fig:case}
\end{wrapfigure}
Task completion is judged against the original instruction, while simulated users rate how well the outcome meets their requests. At delivery, mean satisfaction is 4.04 for the original request and 3.10--3.59 across persona groups (Figure~\ref{fig:domain-persona}). Yet completing the task does not guarantee satisfaction. Among 646 fully successful persona cells, only 62.7\% receive the top rating of 5 and 25.9\% receive 3 or lower, compared with 84.1\% and 9.5\% for fully successful original requests.

Conversely, users {often} perceive failed tasks as complete. Among 800 failed action cells under personas, 56\% are judged done at delivery, falling to 27\% after site-state verification. Two factors contribute to this mismatch. First, 138 of the 343 action loss cells involve correct stops requested by users, 62\% of which are judged done at delivery. Second, agents sometimes misreport outcomes: final replies are inconsistent with the outcome in 45\% of information loss cells and 26\% of action loss cells. Among the remaining 205 action loss cells, misreporting is associated with higher satisfaction (3.47 vs. 3.13) and {the share judged done} (55\% vs. 41\%; Table~\ref{tab:stops}).

Together, these findings show that task completion and user satisfaction diverge in both directions: fully successful tasks can leave users dissatisfied, while failed tasks {can} seem done because agents correctly stop or misrepresent their outcomes.

\paragraph{Case Study: Good Score, Unhappy User}
\label{sec:res-case}
Figure~\ref{fig:case} follows one \textsc{practical} user who asks for a simple dinner from a store's buy-one-get-one offers. The agent {engages with} the request but suggests pork chops or steak with bacon or sausage, so it meets three of four rubric items, and the user's rating of 2/5 names this list of options.  {A partial score can thus hide how far an answer falls from what the user asked.}

\subsection{Where the Drop Comes From}
\label{sec:res-form}

\begin{wrapfigure}{R}{0.48\textwidth}
\centering
\includegraphics[width=\linewidth]{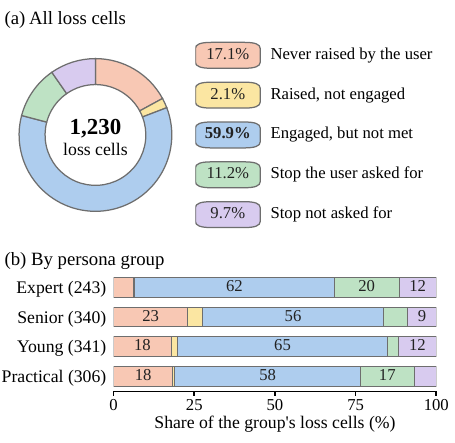}
\caption{Primary cause of each loss cell, nine models pooled. Each cell has one cause. (a) All 1,230 loss cells. (b) Each persona group, with its number of loss cells.}
\label{fig:loss-causes}
\end{wrapfigure}

The previous subsection shows how users perceive the results. To trace how the drop arises, we take the 1,230 loss cells, the persona cells that score below the original request on the same task and backbone, and give each one a primary cause. A cell that ends without the final submission is a stop, either one the user asked for by requesting a draft, a review, or to finish the action themselves, or one the agent made on its own. For every other cell we look at the requirements it lost and take the earliest break between the user and the outcome. {The user may never have raised the requirement, the agent may not have engaged with a requirement the user raised, or the agent may have engaged with it and still not met it.} Figure~\ref{fig:loss-causes} shows the shares.

Most loss cells arise from how agents act on what users say. In 59.9\% of loss cells the agent {engaged with} the requirements the user {raised} and still missed at least one of them. Only 17.1\% lose points solely on requirements the user never {raised}, and 2.1\% on a {raised} requirement that the agent never {engaged with}. The remaining 20.9\% end before the final step. In more than half of these, 11.2\% of all loss cells, the user had asked for a draft or a review first, so stopping is what the user wanted even though the checker scores it as a failure.

The mix of causes differs by group. {Requirements the user never raised} account for 22.9\% of the losses of \textsc{senior}s against 6.2\% for \textsc{expert}s, so non-experts leave more unsaid. Stops that the user asked for account for 20.2\% of \textsc{expert} losses and 16.7\% of \textsc{practical} losses, against 3.5\% for \textsc{young} users. {This is consistent with \textsc{expert} requests reaching the lowest task completion in professional and tying for the lowest in productivity, the two action-only domains of Figure~\ref{fig:domain-persona}.}

This analysis reveals that most loss cells arise after the agent has heard the requirement. Users leave some requirements unsaid, most of all \textsc{senior} users, but the larger share of loss cells comes from {raised} requirements that the agent {engages with} and still gets wrong. The rest is largely stops that the user asked for, where following the user conflicts with the original checker.

\subsection{Interventions}
\label{sec:res-profile}

The analyses so far are observational. To test whether more information about the user, or more interaction with it, recovers the loss, we rerun GPT-6 Astra with one change at a time. Both runs reuse the opening requests of the main experiment and keep the tasks, tools, budgets, persona agents, and judges fixed. The profile-visible run reruns the 296 persona cells with the user's profile added to the agent's input. The forced-question run reruns all 370 cells, including the original requests, and requires the agent to ask the user at least one question before it acts. Appendix~\ref{app:res-ablation} gives the full results.

\textbf{Intervention 1: Showing the agent the profile.}
With the profile, answers are longer for \textsc{expert} users and shorter for \textsc{senior} users. Information-task completion decreases by $0.027$ (95\% CI $[-0.051,-0.001]$), with the largest estimated decline among \textsc{senior} users ($\Delta=-0.070$, 95\% CI $[-0.115,-0.025]$). Action-task completion changes by $+0.028$ (95\% CI $[-0.007,+0.069]$). Profiles alone do not recover the loss.

\textbf{Intervention 2: Forcing the agent to ask.}
\label{sec:res-force}
When the agent must ask before it acts, every cell contains a question, yet task completion on persona requests does not recover, moving from 0.434 to 0.421. On action tasks with the original request the extra question even lowers task completion from 0.583 to 0.472, because the persona agent answers with constraints absent from the instruction. Questions also rarely reach what the user left unsaid: only 24\% of the 609 questions in the main experiment target an omitted requirement.

\section{Related Work}
\label{sec:related}

\paragraph{Computer-use agent benchmarks.}
WebArena \citep{webarena} checks 812 web tasks with fixed instructions by program. REAL \citep{real} checks the final site state on replicas of 11 websites. Odysseys \citep{odysseys} scores 200 long-horizon web tasks with per-requirement rubrics. Other benchmarks cover shopping, real and multimodal websites, and navigation guided by dialogue~\citep{webshop, mind2web, visualwebarena, weblinx}. OSWorld 2.0 \citep{osworld2} adds a simulated user that the agent can query. MyPCBench \citep{mypcbench} fills a desktop with one persona's data but keeps the instruction fixed. None of them poses the same task for different users, as Table~\ref{tab:related} shows.

\begin{wraptable}{R}{0.48\textwidth}
\centering
\caption{Comparison with related benchmarks.}
\label{tab:related}

\footnotesize
\setlength{\tabcolsep}{0pt}
\begin{tabular*}{\linewidth}{@{\extracolsep{\fill}}l c|c|c|c@{}}
\toprule
\textbf{Benchmark} & \tablehead{\textbf{Computer}\\\textbf{use}} & \tablehead{\textbf{Same task,}\\\textbf{many users}} & \tablehead{\textbf{Can ask}\\\textbf{the user}} & \tablehead{\textbf{Judged}\\\textbf{by user}} \\
\midrule
WebArena & \cmark & \xmark & \xmark & \xmark \\
OSWorld 2.0 & \cmark & \xmark & \cmark & \xmark \\
MyPCBench & \cmark & \xmark & \xmark & \xmark \\
PersonalWAB & \xmark & \xmark & \cmark & \xmark \\
Persona2Web & \cmark & \xmark & \xmark & \xmark \\
PrefDisco & \xmark & \cmark & \cmark & \xmark \\
\midrule
\textbf{PersonaCUABench} & \cmark & \cmark & \cmark & \cmark \\
\bottomrule
\end{tabular*}
\end{wraptable}
\paragraph{Personalized agents.}
PersonalWAB \citep{personalwab} gives the agent the profiles and behavior histories of 1,000 users for personalized search and recommendation. Persona2Web \citep{persona2web} uses the browsing histories of 50 users to resolve ambiguous queries. Benchmarks for language models likewise condition on profiles, histories, or stated preferences~\citep{lamp, prefeval, personalens}, and model answers get worse for some groups of users~\citep{targetedunderperf}. PrefDisco \citep{li2025prefdisco} assigns ten personas to each reasoning problem and finds that models ask only 1.42 questions on average within five turns. Other benchmarks test whether agents ask about vague or underspecified instructions~\citep{tellmemore, clamber, askbeforeplan, ambibench}, including those of older adults~\citep{elderbench}. These works study how an agent obtains user information. We record for each requirement whether it was raised, engaged, and met, and find that most loss cells arise after the agent has the information.

\paragraph{LLM user simulation.}
$\tau$-bench \citep{yao2024taubench} pairs a tool-using agent with an LLM-simulated customer, and other benchmarks simulate users who give feedback, reveal preferences, or spread a request over turns~\citep{mint, userbench, lostinconv}. \citet{zhou2026sim2real} replace the simulated users of $\tau$-bench with 451 people and find the simulated ones more verbose, more polite, and more complete at the start, and simulated users can misjudge agents for speakers of some dialects~\citep{lostinsim}. Our persona agents, like those of PersonaGym~\citep{personagym}, are conditioned on a profile, here 37 attributes from MatrAIx \citep{li2026matraixsimulatingworld83}, and state the request and rate the outcome in the first person, which exposes the start and end of the service.

\section{Conclusion}
\label{sec:conclusion}

PersonaCUABench poses the same computer-use tasks for different users, and persona requests lower task completion from 0.689 to 0.441 across nine backbones. Most loss cells arise after the agent has engaged with what the user raised, and another 11.2\% are stops the user asked for that the checker scores as failures. A fixed specification thus conflates how well the agent executes with how far the user's request departs from it. On GPT-6 Astra, neither showing the profile nor requiring a question closes the gap: the bottleneck is carrying what the user said through a long sequence of actions. As completion and satisfaction also diverge, personal CUAs should be judged against the user's request and from the user's side, with simulated users validated against real ones.

\section*{Limitations}

PersonaCUABench spans 11 domains drawn from everyday and work settings and evaluates nine general-purpose backbones in a shared browser harness. Its tasks focus on finding information and taking actions on websites, leaving desktop applications, standalone chat, and coding workflows outside its scope.  Each episode begins with a single request and focuses on one task, whereas personal agents in real-world use can accumulate knowledge about their users across interactions. Future extensions could evaluate repeated interactions with persistent user memory and extend the intervention study to additional backbones.

\section*{Ethical Statement}
PersonaCUABench uses constructed personas and LLM-simulated users. Information tasks access live websites, while action tasks run on sandboxed replicas to avoid real-world transactions or communications. We acknowledge that persona construction may encode stereotypes and cannot capture the full diversity within demographic groups. Accordingly, performance differences should not be generalized to real populations, and simulated satisfaction ratings cannot replace feedback from real users. Our benchmark aims to improve user-aligned evaluation and highlights the importance of respecting user intent and obtaining authorization before consequential actions.

\bibliographystyle{plainnat}
\bibliography{custom}

\clearpage
\appendix
\etocdepthtag.toc{appendix}
\appendixcontents

\section{PersonaCUABench Construction}
\label{app:construction}
This appendix details the materials of \S\ref{sec:construction}: how the tasks were selected, the persona attributes, the persona agent's prompt, and how faithfully the persona agent follows its profile.

\subsection{Task Curation}
\label{app:tasks}
\paragraph{Selection.} Odysseys contains 200 live-web tasks and REAL 112 tasks on replicas of 11 websites. From the candidates we considered in the two sources, we dropped 20 tasks: 12 live-web tasks whose site blocked automated access or did not load, three REAL tasks on the eleventh replica (Omnizon), whose host was unreachable from our network, two pure navigation tasks that leave no room for preferences, two duplicates of kept tasks, and one task that requires watching a video. Our action tasks therefore use ten of the eleven REAL websites. By the source benchmarks' own labels, the information tasks are 18 easy and 20 medium, and the action tasks 35 medium and one hard.

\paragraph{Requirements of REAL tasks.} A REAL task comes with one instruction and one pass/fail checker. To analyze intent on these tasks, we derive the requirements that each checker tests from the task manifest and the checker's code (e.g., ``the order includes a pizza'' and ``the order total is under \$30''), phrase them as what the user wants, and review them by hand. These 107 requirements serve the intent measures of \S\ref{sec:eval}.

\paragraph{Domains.} Each task belongs to one of 11 domains, assigned through its website. A domain can group several websites. Travel, for instance, combines five Odysseys tasks with REAL airline, lodging, and ride-hailing replicas, whereas professional and productivity contain only REAL tasks. The tasks span two scenarios, 40 life and 34 work tasks. Figure~\ref{fig:overview} (top) shows how the 74 tasks spread over the 11 domains.

\subsection{Persona Attributes}
\label{app:attributes}
Every persona carries the 32 general attributes of Table~\ref{tab:attr-general} and the five attributes of its domain in Table~\ref{tab:attr-domain}, all taken from the MatrAIx schema~\citep{li2026matraixsimulatingworld83}. The first five general attributes define the persona groups (Table~\ref{tab:roles}), and technology savviness, together with seniority and the domain-specific skills, reflects expertise. Five values in four personas differ from Table~\ref{tab:roles}: the travel \textsc{senior} persona has a household income below \$25k; the travel \textsc{practical} persona, a recent immigrant, has an income of \$25k--50k and entry-level seniority; and the art and professional \textsc{practical} personas are at the senior level.

\begin{table*}[t]
\centering
\small
\caption{The 32 general attributes shared by all personas, grouped by the question they answer (10, 7, 7, and 8 attributes).}
\label{tab:attr-general}
\setlength{\tabcolsep}{3pt}
\renewcommand{\arraystretch}{1.05}
\begin{tabular}{@{}>{\raggedright\arraybackslash}p{86pt}
                >{\raggedright\arraybackslash}p{99pt}
                >{\raggedright\arraybackslash}p{126pt}
                >{\raggedright\arraybackslash}p{124pt}@{}}
\toprule
\textbf{Who they are} & \textbf{Can they use the result} & \textbf{What they ask for and expect} & \textbf{Which trade-offs they accept} \\
\midrule
\hangindent=0.8em\hangafter=1 age bracket\par\hangindent=0.8em\hangafter=1 highest education\par\hangindent=0.8em\hangafter=1 household income\par\hangindent=0.8em\hangafter=1 time pressure\par\hangindent=0.8em\hangafter=1 seniority\par\hangindent=0.8em\hangafter=1 technology savviness\par\hangindent=0.8em\hangafter=1 region\par\hangindent=0.8em\hangafter=1 urbanicity\par\hangindent=0.8em\hangafter=1 socioeconomic band\par\hangindent=0.8em\hangafter=1 geographic mobility &
\hangindent=0.8em\hangafter=1 interest in technology\par\hangindent=0.8em\hangafter=1 attitude to new technology\par\hangindent=0.8em\hangafter=1 device context\par\hangindent=0.8em\hangafter=1 mobility\par\hangindent=0.8em\hangafter=1 vision\par\hangindent=0.8em\hangafter=1 assistive technology\par\hangindent=0.8em\hangafter=1 accessibility needs &
\hangindent=0.8em\hangafter=1 detail orientation\par\hangindent=0.8em\hangafter=1 patience\par\hangindent=0.8em\hangafter=1 attention span\par\hangindent=0.8em\hangafter=1 big picture vs.\ detail\par\hangindent=0.8em\hangafter=1 brief vs.\ full detail\par\hangindent=0.8em\hangafter=1 planning horizon\par\hangindent=0.8em\hangafter=1 planned vs.\ spontaneous &
\hangindent=0.8em\hangafter=1 budgeting skill\par\hangindent=0.8em\hangafter=1 frugality\par\hangindent=0.8em\hangafter=1 economic motivation\par\hangindent=0.8em\hangafter=1 budget tracking\par\hangindent=0.8em\hangafter=1 save vs.\ spend preference\par\hangindent=0.8em\hangafter=1 risk tolerance\par\hangindent=0.8em\hangafter=1 financial risk tolerance (DOSPERT)\par\hangindent=0.8em\hangafter=1 risk framing \\
\bottomrule
\end{tabular}
\end{table*}

\begin{table*}[t]
\centering
\small
\caption{The five domain-specific attributes of each domain, grouped by type. \emph{Knowledge} covers familiarity with a field and subject knowledge; \emph{Other} covers values, styles, and habits. A dash marks a type that a domain does not use.}
\label{tab:attr-domain}
\setlength{\tabcolsep}{5pt}
\renewcommand{\arraystretch}{1.05}
\begin{tabular}{@{}>{\raggedright\arraybackslash}p{0.10\textwidth}
                >{\raggedright\arraybackslash}p{0.17\textwidth}
                >{\raggedright\arraybackslash}p{0.14\textwidth}
                >{\raggedright\arraybackslash}p{0.18\textwidth}
                >{\raggedright\arraybackslash}p{0.12\textwidth}
                >{\raggedright\arraybackslash}p{0.18\textwidth}@{}}
\toprule
\textbf{Domain} & \textbf{Industry} & \textbf{Skills} & \textbf{Knowledge} & \textbf{Interests} & \textbf{Other} \\
\midrule
\mbox{Art} & \hangindent=0.8em\hangafter=1 fine art & \hangindent=0.8em\hangafter=1 drawing\par\hangindent=0.8em\hangafter=1 design thinking & \hangindent=0.8em\hangafter=1 art history & \textcolor{black!35}{--} & \hangindent=0.8em\hangafter=1 appreciation of beauty \\
\arrayrulecolor{black!20}\specialrule{0.4pt}{2.5pt}{2.5pt}\arrayrulecolor{black}
\mbox{Community} & \hangindent=0.8em\hangafter=1 nonprofit\par\hangindent=0.8em\hangafter=1 government & \hangindent=0.8em\hangafter=1 facilitation & \textcolor{black!35}{--} & \hangindent=0.8em\hangafter=1 volunteering & \hangindent=0.8em\hangafter=1 value: community \\
\arrayrulecolor{black!20}\specialrule{0.4pt}{2.5pt}{2.5pt}\arrayrulecolor{black}
\mbox{E-commerce} & \hangindent=0.8em\hangafter=1 e-commerce\par\hangindent=0.8em\hangafter=1 retail & \hangindent=0.8em\hangafter=1 negotiation & \textcolor{black!35}{--} & \textcolor{black!35}{--} & \hangindent=0.8em\hangafter=1 shopping style\par\hangindent=0.8em\hangafter=1 brand loyalty \\
\arrayrulecolor{black!20}\specialrule{0.4pt}{2.5pt}{2.5pt}\arrayrulecolor{black}
\mbox{Food} & \hangindent=0.8em\hangafter=1 food and beverage\par\hangindent=0.8em\hangafter=1 hospitality & \hangindent=0.8em\hangafter=1 cooking & \textcolor{black!35}{--} & \hangindent=0.8em\hangafter=1 cooking & \hangindent=0.8em\hangafter=1 diet type \\
\arrayrulecolor{black!20}\specialrule{0.4pt}{2.5pt}{2.5pt}\arrayrulecolor{black}
\mbox{Health} & \hangindent=0.8em\hangafter=1 healthcare\par\hangindent=0.8em\hangafter=1 fitness and wellness & \textcolor{black!35}{--} & \hangindent=0.8em\hangafter=1 nutrition & \hangindent=0.8em\hangafter=1 fitness & \hangindent=0.8em\hangafter=1 exercise frequency \\
\arrayrulecolor{black!20}\specialrule{0.4pt}{2.5pt}{2.5pt}\arrayrulecolor{black}
\mbox{Lifestyle} & \hangindent=0.8em\hangafter=1 apparel\par\hangindent=0.8em\hangafter=1 real estate & \textcolor{black!35}{--} & \textcolor{black!35}{--} & \hangindent=0.8em\hangafter=1 interior design\par\hangindent=0.8em\hangafter=1 fashion & \hangindent=0.8em\hangafter=1 fashion sense \\
\arrayrulecolor{black!20}\specialrule{0.4pt}{2.5pt}{2.5pt}\arrayrulecolor{black}
\mbox{Productivity} & \textcolor{black!35}{--} & \hangindent=0.8em\hangafter=1 time management\par\hangindent=0.8em\hangafter=1 writing & \hangindent=0.8em\hangafter=1 organizational psychology & \hangindent=0.8em\hangafter=1 productivity & \hangindent=0.8em\hangafter=1 inbox-zero discipline \\
\arrayrulecolor{black!20}\specialrule{0.4pt}{2.5pt}{2.5pt}\arrayrulecolor{black}
\mbox{Professional} & \hangindent=0.8em\hangafter=1 consulting & \hangindent=0.8em\hangafter=1 networking\par\hangindent=0.8em\hangafter=1 leadership & \textcolor{black!35}{--} & \textcolor{black!35}{--} & \hangindent=0.8em\hangafter=1 value: career success\par\hangindent=0.8em\hangafter=1 career path shape \\
\arrayrulecolor{black!20}\specialrule{0.4pt}{2.5pt}{2.5pt}\arrayrulecolor{black}
\mbox{Science} & \hangindent=0.8em\hangafter=1 education & \hangindent=0.8em\hangafter=1 research\par\hangindent=0.8em\hangafter=1 mathematics & \hangindent=0.8em\hangafter=1 statistics & \hangindent=0.8em\hangafter=1 science & \textcolor{black!35}{--} \\
\arrayrulecolor{black!20}\specialrule{0.4pt}{2.5pt}{2.5pt}\arrayrulecolor{black}
\mbox{Technology} & \hangindent=0.8em\hangafter=1 consumer electronics & \hangindent=0.8em\hangafter=1 coding\par\hangindent=0.8em\hangafter=1 debugging & \hangindent=0.8em\hangafter=1 computer networking\par\hangindent=0.8em\hangafter=1 computer science & \textcolor{black!35}{--} & \textcolor{black!35}{--} \\
\arrayrulecolor{black!20}\specialrule{0.4pt}{2.5pt}{2.5pt}\arrayrulecolor{black}
\mbox{Travel} & \hangindent=0.8em\hangafter=1 travel and tourism\par\hangindent=0.8em\hangafter=1 transportation & \textcolor{black!35}{--} & \textcolor{black!35}{--} & \hangindent=0.8em\hangafter=1 travel & \hangindent=0.8em\hangafter=1 travel breadth\par\hangindent=0.8em\hangafter=1 travel frequency \\
\bottomrule
\end{tabular}
\end{table*}

\subsection{Persona Agent Prompt}
\label{app:persona-prompt}
The persona agent's system message joins a role-play preamble, the rendered profile $\pi(\mathbf{v}_p)$, and the rules below, in which \texttt{\{goal\}} is replaced by the original instruction $x_t$. It then receives ``Write the message you send the assistant. Only the message itself.'' to produce $u_0$, and each clarification question as ``The assistant asks you: $q_j$. Reply as yourself.'' Under the original request, the agent receives $x_t$ verbatim, and questions are answered under the preamble ``You are a person.'' with the same rules and no profile.

\clearpage
\begin{figure*}[p]
\centering
\begin{minipage}[b]{\textwidth}
\raggedright
{\large\bfseries Persona agent system prompt}\par
{\small\emph{Colored card titles and icons are editorial annotations, not part of the prompt. The prompt text and its reading order are preserved below.}\par}
\input{persona_agent_prompt}
\end{minipage}
\end{figure*}

\clearpage
\subsection{Persona Fidelity}
\label{app:fidelity}

\begin{figure}[t]
\centering
\includegraphics[width=0.6\linewidth]{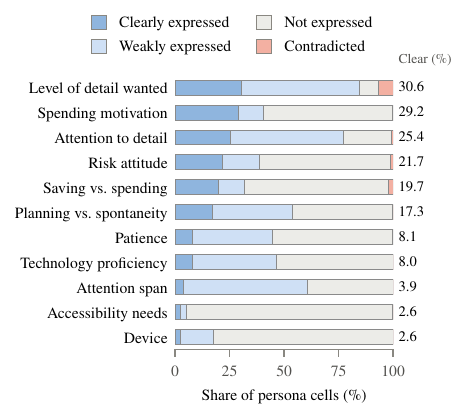}
\caption{How often each profile dimension appears in the persona's utterances (2,664 persona cells over nine models). The right column gives the share of cells in which the dimension is clearly expressed.}
\label{fig:fidelity}
\end{figure}
A judge (GPT-5.6-terra) reads only the persona agent's utterances and rates each of 11 profile dimensions as clearly expressed, weakly expressed, not expressed, or contradicted (Figure~\ref{fig:fidelity}, Table~\ref{tab:fidelity}). Contradictions do not drive the drop: 11.1\% of cells contain at least one, and removing them changes the mean $\Delta$ from $-0.248$ to $-0.249$. Of the 329 contradictions across models, 168 are \textsc{expert} personas whose profile specifies that they want every detail but who request a brief answer. The profile is, however, expressed only in part: on average 15.4\% of the dimensions are clearly expressed, and 66.8\% of cells clearly express at least one. Accessibility and device are rated as not expressed in 95\% and 82\% of judgments.

\begin{table*}[t]
\centering
\small
\setlength{\tabcolsep}{4.5pt}
\caption{Persona fidelity per model. Clear is the share of the 11 profile dimensions that the persona's utterances state clearly. Contradiction is the share of cells with at least one utterance that contradicts the profile. The $\Delta$ column gives the overall $\Delta$ on all cells and with those cells excluded. Accessibility and Device give the share of cells in which the accessibility or device dimension is clearly expressed.}
\label{tab:fidelity}
\begin{tabular}{@{}l c c c c c@{}}
\toprule
\textbf{Model} & \textbf{Clear (\%)} & \textbf{Contradiction (\%)} & \textbf{$\Delta$: all $\to$ excluded} & \textbf{Accessibility (\%)} & \textbf{Device (\%)} \\
\midrule
\vendoricon{minimax}MiniMax M3          & 16.1 & 11.1 & $-0.305 \to -0.317$ & 1.7 & 2.0 \\
\vendoricon{kimi}Kimi K3             & 14.5 & 9.1  & $-0.284 \to -0.285$ & 2.4 & 1.4 \\
\vendoricon{muse}Muse Glimmer 30B    & 15.0 & 11.8 & $-0.194 \to -0.191$ & 3.4 & 4.7 \\
\vendoricon{grok}Grok 4.6            & 13.8 & 8.8  & $-0.190 \to -0.194$ & 3.7 & 2.7 \\
\vendoricon{qwen}Qwen 3.8 27B        & 15.0 & 12.5 & $-0.233 \to -0.225$ & 3.4 & 4.4 \\
\vendoricon{gemini}Gemini 3.8 Flash    & 13.5 & 7.8  & $-0.311 \to -0.313$ & 2.0 & 1.7 \\
\vendoricon{gpt}GPT-6 Astra         & 17.8 & 13.5 & $-0.241 \to -0.255$ & 2.4 & 2.0 \\
\vendoricon{deepseek}DeepSeek V4.1 Flash & 14.1 & 12.2 & $-0.259 \to -0.244$ & 1.4 & 2.7 \\
\vendoricon{claude}Claude Opus 5.5     & 18.6 & 13.2 & $-0.212 \to -0.213$ & 3.4 & 2.0 \\
\midrule
\textbf{Mean}       & 15.4 & 11.1 & $-0.248 \to -0.249$ & 2.6 & 2.6 \\
\bottomrule
\end{tabular}

\end{table*}

\section{Experimental Setup and Evaluation}
\label{app:setup-eval}
This appendix details how cells are run and scored (\S\ref{sec:eval} and \S\ref{sec:setup}): the harness, the full definitions of the measures, the API cost, and the reported agreement of the judges with P1 and P2.

\subsection{Harness}
\label{app:harness}
All backbones run in the same Chrome harness with the same nine tools: \texttt{open\_tab}, \texttt{goto}, \texttt{read\_page}, \texttt{click}, \texttt{type\_text}, \texttt{list\_tabs}, \texttt{close\_tab}, \texttt{ask\_user}, and \texttt{finish}. The observation is page text. \texttt{read\_page} returns the visible text followed by the page's interactive fields and controls, each with a handle the agent can target. Screenshots are stored at every step for auditing but never enter the agent's context. For REAL tasks the harness loads the prescribed initial state $s_0$ before the episode and reads the final state $s_T$ afterward. A further question beyond the budget of $Q=5$ is refused, and the agent is told to decide for itself. The episode ends when the agent calls \texttt{finish}. At the turn limit it receives one more prompt to summarize its findings. Appendix~\ref{app:cost} reports the API cost.

\subsection{Evaluation Details}
\label{app:eval}
This appendix gives the full definitions behind the measures of \S\ref{sec:eval} and \S\ref{sec:protocol}, covering task completion, intent, and user satisfaction.

\paragraph{Task completion.} The evaluation procedure $C_t$ takes one of two forms. An LLM rubric judge decides each $m_i$ from the agent's final answer $y$, its action log, and the tabs $\Omega$ left open, applying each requirement's verification condition: a requirement that a page be left open is met only if that page is among the open tabs, a reported figure counts only if the agent visited a page that shows it, and a requirement naming a specific site is met only by that site. The judge does not see the dialogue, so $m_i$ depends only on the outcome. A programmatic checker instead reads the site's final state $s_T$ and returns one verdict $C_t(s_T)$, which sets $m_i$ for every requirement of the task. Odysseys tasks use the rubric judge and REAL tasks the state checker.

\paragraph{Intent.} Raised, $\rho_i = J_{\text{raise}}(r_i, \delta_U) \in \{0,1\}$, records whether the persona stated $r_i$ directly, by clear implication, or by confirming it when asked. It is computed from the persona's utterances $\delta_U$ alone, so it depends on the agent only through the questions the agent asks. Engaged, $\varepsilon_i = J_{\text{eng}}(r_i, \tau) \in \{0,1\}$, records whether the agent took $r_i$ up at all, by pursuing it, addressing it, or asking about it, irrespective of correctness. Together they place each requirement in one of four outcomes (Table~\ref{tab:intent}). For REAL tasks, $\rho_i$ and $\varepsilon_i$ are computed over the requirements extracted from each checker (Appendix~\ref{app:tasks}).

From these judgements we compute the coverage of a request, the share of its requirements that the user raised:
\begin{equation}
\text{coverage} = \frac{1}{n_t}\sum_{i=1}^{n_t} \rho_i.
\label{eq:intent}
\end{equation}
Read with $m_i$, the three judgements locate where service breaks for a requirement: it was not raised, it was raised but not engaged, or it was engaged but not met. Engaged records that the agent acted on $r_i$, not that it read $r_i$ correctly, so a requirement that is engaged but not met can reflect a misreading as well as a failed execution.

\begin{wraptable}{R}{0.48\textwidth}
\centering
\small
\caption{Intent outcomes for a requirement $r_i$, by whether the user raised it ($\rho_i$) and whether the agent engaged with it ($\varepsilon_i$).}
\label{tab:intent}
\begin{tabular}{@{}lcc@{}}
\toprule
& $\varepsilon_i = 1$ & $\varepsilon_i = 0$ \\
\midrule
$\rho_i = 1$ & engaged & dropped \\
$\rho_i = 0$ & engaged unprompted & unsaid \\
\bottomrule
\end{tabular}
\end{wraptable}
\paragraph{User satisfaction.} Satisfaction is rated by the persona agent $U$ that made the request, so each rollout is assessed by the person it was meant to serve. After the episode, $U$ is rebuilt with its profile $\pi(\mathbf{v}_p)$, its goal $x_t$, and the exact conversation it had with the agent, and is shown what a user would see: the agent's final answer $y$ and the tabs $\Omega$ left open. Staying in character, it returns a satisfaction score from 1 (very dissatisfied) to 5 (very satisfied), whether it believes the agent did what it asked (yes, no, or unsure), and a one-sentence reason. For action tasks it is asked a second time after verification, when it is also shown what changed on the site. For the original request, $U$ rates without a profile. Only the latest scored run of each cell is rated. Because $U$ is rebuilt with $x_t$, its rating can draw on the original instruction as well as on what it asked for. Appendix~\ref{app:res-report} discusses what this means for stops the user asked for. \textcolor{black}{Appendix~\ref{app:sat-check} compares both rating stages with human satisfaction annotations on the 22 selected tasks.}

We also ran an external judge, used only to check the user's rating in Appendix~\ref{app:res-report}. The external judge, GPT-5.6-terra, sees the persona's profile, the original instruction $x_t$ (marked as unseen by the agent), the full dialogue, the agent's actions, its final answer, and the open tabs. It judges the agent against what this person said and who they are, and scores five dimensions on a 1--5 scale (or n/a when the rollout gives no evidence):
\begin{itemize}
\item \textbf{P1 Asked-for}: whether the agent delivered what this person stated, clearly implied, or confirmed when asked.
\item \textbf{P2 Readable}: whether this person can read and understand the answer, given their vocabulary, technology familiarity, attention span, and constraints such as low vision or a phone.
\item \textbf{P3 Actionable}: whether this person can act on the answer without help they lack, with concrete next steps, prices, times, and links, and with the pages they asked for left open.
\item \textbf{P4 Gap handling}: whether the agent noticed what this person left unsaid, through a question, a stated assumption, or a flagged risk.
\item \textbf{P5 Overall}: whether this person would say they were well served, judged as a whole.
\end{itemize}
The judge also lists explicit requests that went unmet and unstated needs that the agent missed.

\subsection{API Cost}
\label{app:cost}
For each cell we log tokens (including cached and reasoning tokens), monetary cost, wall-clock time, and step count, with judging cost kept separate from agent cost. Table~\ref{tab:cost} reports the API cost over the cells in the paper, using the latest scored run of each cell. We cost every agent backbone at its list price per million input and output tokens, without cache discounts, so that the backbones are comparable. Providers bill cached input at a discount, and 68\% of all agent input tokens are cached, so the billed cost is lower than the list cost. Over the eight backbones with a list price, the agent costs 4,737 USD, or 1.60 USD per cell. The cost per cell ranges from 0.26 USD for DeepSeek V4.1 Flash to 3.26 USD for Kimi K3 and follows the price per token more than the number of steps. Muse Glimmer 30B has no list price, so we report its tokens only.

The persona agents and all automated judges run on GPT-5.6-terra, for which the run records contain tokens but no price. We estimate their cost at its list price of 2.00 USD per million input tokens and 12.00 USD per million output tokens, again without cache discounts. This gives about 202 USD over all nine backbones, or 17 to 26 USD per backbone, an order of magnitude below the agents. The estimate covers the persona agents' replies to questions, the rubric, intent, external, and fidelity judges, and the satisfaction ratings. It leaves out the generation of the opening requests.

\begin{table*}[t]
\centering
\small
\setlength{\tabcolsep}{0pt}
\caption{API cost per backbone at list prices without cache discounts. The persona agents and judges run on GPT-5.6-terra and are estimated at 2.00 / 12.00 USD per million input / output tokens. $^{\dagger}$Eight backbones with a list price.}
\label{tab:cost}
\begin{tabular*}{\textwidth}{@{\extracolsep{\fill}}l c r r r r r r@{}}
\toprule
\multicolumn{1}{c}{\tablehead{\textbf{Model}}} &
\multicolumn{1}{c}{\tablehead{\textbf{List price,}\\\textbf{input / output}\\\textbf{(USD per million)}}} &
\multicolumn{1}{c}{\tablehead{\textbf{Input}\\\textbf{tokens}\\\textbf{(millions)}}} &
\multicolumn{1}{c}{\tablehead{\textbf{Output}\\\textbf{tokens}\\\textbf{(millions)}}} &
\multicolumn{1}{c}{\tablehead{\textbf{Mean}\\\textbf{steps}}} &
\multicolumn{1}{c}{\tablehead{\textbf{Agent}\\\textbf{cost}\\\textbf{(USD)}}} &
\multicolumn{1}{c}{\tablehead{\textbf{Agent cost}\\\textbf{per cell}\\\textbf{(USD)}}} &
\multicolumn{1}{c}{\tablehead{\textbf{Persona agent}\\\textbf{and judges,}\\\textbf{estimated (USD)}}} \\
\midrule
\vendoricon{minimax}MiniMax M3           & 0.30 / 1.20   & 332.1 & 3.39 & 43.1 & 104   & 0.28 & 18 \\
\vendoricon{kimi}Kimi K3                 & 3.00 / 15.00  & 372.9 & 5.75 & 35.8 & 1,205 & 3.26 & 17 \\
\vendoricon{muse}Muse Glimmer 30B        & --            & 293.6 & 2.49 & 41.6 & --    & --   & 25 \\
\vendoricon{grok}Grok 4.6                & 2.00 / 6.00   & 397.4 & 2.55 & 42.9 & 810   & 2.19 & 26 \\
\vendoricon{qwen}Qwen 3.8 27B            & 0.42 / 3.00   & 404.0 & 5.69 & 35.1 & 187   & 0.50 & 25 \\
\vendoricon{gemini}Gemini 3.8 Flash      & 0.75 / 3.75   & 384.6 & 3.02 & 36.8 & 300   & 0.82 & 21 \\
\vendoricon{gpt}GPT-6 Astra              & 10.00 / 50.00 & 110.6 & 0.66 & 28.4 & 1,140 & 3.08 & 22 \\
\vendoricon{deepseek}DeepSeek V4.1 Flash & 0.30 / 1.20   & 303.4 & 3.67 & 39.6 & 95    & 0.26 & 25 \\
\vendoricon{claude}Claude Opus 5.5       & 4.00 / 20.00  & 214.3 & 1.98 & 25.1 & 897   & 2.42 & 24 \\
\midrule
\textbf{Total}                           &               & 2,813.0 & 29.20 & 36.5 & 4,737$^{\dagger}$ & 1.60$^{\dagger}$ & 202 \\
\bottomrule
\end{tabular*}

\end{table*}

\subsection{Human Annotation}
\label{app:human}
\begin{wraptable}{R}{0.48\textwidth}
\centering
\footnotesize
\setlength{\tabcolsep}{2.5pt}
\caption{Agreement of human annotators P1 and P2 with the GPT-5.6-terra judge on 22 representative tasks (two per domain), as percent agreement with Cohen's $\kappa$ in parentheses. Items count the labels that GPT-5.6-terra gives. For each profile dimension, Clear asks whether it is clearly expressed, and Level asks whether it is clearly, weakly, or not expressed.}
\label{tab:human}
\begin{tabular*}{\linewidth}{@{\extracolsep{\fill}}l r c c@{}}
\toprule
\textbf{Judgment} & \textbf{Items} & \textbf{P1} & \textbf{P2} \\
\midrule
\multicolumn{4}{c}{\textit{Requirement}} \\
\quad Met & 3,957 & 86.2 (0.72) & 88.2 (0.75) \\
\quad Raised & 3,645 & 88.1 (0.47) & 84.7 (0.20) \\
\quad Engaged & 3,645 & 97.2 (0.55) & 95.9 (0.30) \\
\multicolumn{4}{c}{\textit{Profile dimension}} \\
\quad Clear & 8,712 & 90.7 (0.59) & 88.1 (0.54) \\
\quad Level & 8,712 & 72.8 (0.55) & 64.9 (0.39) \\
\bottomrule
\end{tabular*}
\end{wraptable}
We check the GPT-5.6-terra judgments against annotations from two human annotators, P1 and P2. The annotation set contains 22 tasks, with two representative tasks selected from each of the 11 domains. Each task is run with the four personas and the original request under all nine backbones, yielding 990 outputs with 3,960 requirements and 792 persona-agent dialogues with 11 profile dimensions each (8,712 dimensions). Each annotator reads only the user turns of the dialogue, the agent's answer and trajectory, the rubric, and the profile. They were instructed not to read the GPT-5.6-terra labels, which are used only to compute agreement. For each requirement, they label whether the user raised it, whether the agent engaged with it, and whether it was met (\S\ref{sec:eval}). For each profile dimension, they rate the persona agent's utterances as clearly expressed, weakly expressed, or not expressed, counting a contradiction as not expressed (Appendix~\ref{app:fidelity}). The primary loss cause is not re-annotated. Table~\ref{tab:human} gives the percent agreement with GPT-5.6-terra and Cohen's $\kappa$, counting only items that GPT-5.6-terra labels; GPT-5.6-terra gives no raised or engaged label on the 315 REAL outputs.

Agreement on requirement met is 86.2\% for P1 and 88.2\% for P2 ($\kappa=0.72$ and $0.75$). Raised and engaged agree on 84.7--97.2\% of requirements ($\kappa$ 0.20--0.55); GPT-5.6-terra marks most requirements as raised (82.6\%) and engaged (95.8\%), so chance agreement is high. Whether a profile dimension is clearly expressed agrees on 88.1--90.7\% ($\kappa$ 0.54--0.59).

\subsection[Human Satisfaction Annotation]{\textcolor{black}{Human Satisfaction Annotation}}
\label{app:sat-check}
\begin{wraptable}{R}{0.48\textwidth}
\centering
\footnotesize
\setlength{\tabcolsep}{2.5pt}
\caption{Agreement of human annotators P1 and P2 with the two automated satisfaction ratings on the 22 selected tasks, as exact percent agreement with quadratic-weighted Cohen's $\kappa$ in parentheses. Pairs are matched on backbone, task, persona, and rating stage.}
\label{tab:sat-annotation}
\begin{tabular*}{\linewidth}{@{\extracolsep{\fill}}l c c c@{}}
\toprule
\multicolumn{1}{c}{\textbf{Compared with}} & \textbf{Pairs} & \textbf{P1} & \textbf{P2} \\
\midrule
\multicolumn{4}{c}{\textit{Persona agent's own rating}} \\
\quad At delivery & 987 & 37.5 ($0.37$) & 37.3 ($0.44$) \\
\quad After verification & 305 & 32.1 ($0.33$) & 36.4 ($0.30$) \\
\midrule
\multicolumn{4}{c}{\textit{External judge (P5)}} \\
\quad At delivery & 989 & 48.6 ($0.52$) & 58.2 ($0.60$) \\
\bottomrule
\end{tabular*}
\end{wraptable}
\textcolor{black}{The same two human annotators, P1 and P2, rate satisfaction on the 22 selected tasks. Each annotator rates all 990 outputs, including 792 persona outputs and 198 original-request outputs, on the same 1--5 scale as the persona agent. For delivery ratings, the annotators were instructed to use the user's dialogue, the final answer, the titles and URLs of the open tabs, and the profile when applicable, without consulting the hidden original task or automated judgments. For action tasks, the annotation protocol also asks for a second rating using saved site-state changes.}

We compare these ratings with the two automated satisfaction ratings of Appendix~\ref{app:eval}: the persona agent's own rating, at delivery and after verification, and the external judge's rating, for which we use its overall score P5 at delivery. Ratings are paired on the same backbone, task, persona, and stage, which gives 987 pairs with the persona agent at delivery, 305 after verification, and 989 with the external judge. Table~\ref{tab:sat-annotation} reports exact agreement and quadratic-weighted Cohen's $\kappa$, which accounts for the ordered 1--5 scale.

\textcolor{black}{At delivery, P1 and P2 are within one point of the persona agent's rating on 77.7\% and 79.9\% of outputs, and after verification on 64.6\%. They agree more with the external judge where $\kappa=0.52$ and $0.60$. The annotators and the external judge both rate the outcome from outside the conversation, while the persona agent rates against the full need $x_t$ it was given; its rating is the one that better separates successful from failed outputs (Appendix~\ref{app:res-report}).}

\section{Additional Results}
\label{app:results}
This appendix gives the material behind \S\ref{sec:res-drop} and \S\ref{sec:results}, with one subsection for each part of the main text and in the same order.

\subsection{Overall Results}
\label{app:res-overall}
Table~\ref{tab:main-delta} gives $\Delta$, its 95\% interval, the relative change, and $\Delta$ after excluding cells that contradict the profile, for the comparison in Table~\ref{tab:main}. Across the nine models, the absolute drop grows with the score on original requests, with $r=-0.60$, while the relative drop is unrelated to it, with $r=-0.14$.

\begin{table*}[t]
\centering
\small
\setlength{\tabcolsep}{3pt}
\caption{$\Delta$, 95\% interval, relative change, and $\Delta$ after exclusion, for Table~\ref{tab:main}. $\Delta$ is task completion on the persona request minus that on the original request on the same task, averaged over the four groups. Relative is $\Delta$ divided by the original-request score, in percent. $\Delta$ excluded drops the 11.1\% of cells in which the persona agent contradicts its profile. Every interval lies below zero, and excluding those cells changes the mean $\Delta$ from $-0.248$ to $-0.249$. $\Delta$ pairs each persona cell with the original-request cell of the same task.}
\label{tab:main-delta}
\begin{tabular}{@{}lcccc@{}}
\toprule
\textbf{Model} & $\Delta$ & \textbf{95\% CI} & \textbf{Relative (\%)} & $\Delta$ \textbf{excluded} \\
\midrule
\vendoricon{minimax}MiniMax M3              & $-0.305$ & [$-0.379$, $-0.229$] & $-45$ & $-0.317$ \\
\vendoricon{kimi}Kimi K3                 & $-0.284$ & [$-0.358$, $-0.213$] & $-37$ & $-0.285$ \\
\vendoricon{muse}Muse Glimmer 30B        & $-0.194$ & [$-0.276$, $-0.119$] & $-35$ & $-0.191$ \\
\vendoricon{grok}Grok 4.6                & $-0.190$ & [$-0.256$, $-0.124$] & $-28$ & $-0.194$ \\
\vendoricon{qwen}Qwen 3.8 27B            & $-0.233$ & [$-0.303$, $-0.169$] & $-32$ & $-0.225$ \\
\vendoricon{gemini}Gemini 3.8 Flash        & $-0.311$ & [$-0.381$, $-0.240$] & $-42$ & $-0.313$ \\
\vendoricon{gpt}GPT-6 Astra             & $-0.241$ & [$-0.317$, $-0.167$] & $-36$ & $-0.255$ \\
\vendoricon{deepseek}DeepSeek V4.1 Flash     & $-0.259$ & [$-0.324$, $-0.194$] & $-37$ & $-0.244$ \\
\vendoricon{claude}Claude Opus 5.5         & $-0.212$ & [$-0.283$, $-0.143$] & $-31$ & $-0.213$ \\
\midrule
\textbf{Mean}           & $-0.248$ & [$-0.295$, $-0.202$] & $-36$ & $-0.249$ \\
\bottomrule
\end{tabular}

\end{table*}

\subsection{Generality of the Drop across Domains and Users}
\label{app:full-results}
The performance drop persists across scenarios, domains, and persona groups. All nine models lose on both work and life tasks and across all four persona groups (Table~\ref{tab:groups}), while at least eight of nine models lose in every domain (Figure~\ref{fig:domains}). Work and life tasks lose by similar amounts, $-0.274$ and $-0.226$.

\begin{figure}[t]
\centering
\includegraphics[width=0.65\linewidth]{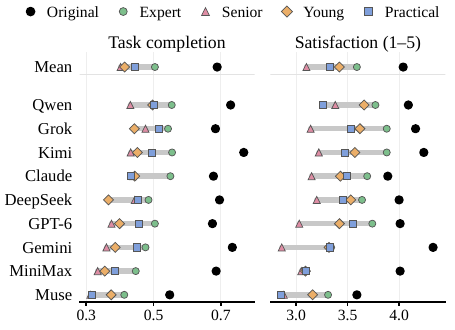}
\caption{Task completion and satisfaction at delivery per model, for the original request and each persona group, over the 74 tasks. The gray bar spans the four persona groups, and the black marker is the original request. Models are sorted by mean persona task completion.}
\label{fig:persona-model}
\end{figure}
The decline also varies across personas: \textsc{expert} loses the least ($-0.185$), followed by \textsc{practical} ($-0.243$) and \textsc{young} ($-0.275$), while \textsc{senior} loses the most ($-0.287$). \S\ref{sec:res-form} examines these differences, and the tables and figures below give the results for each domain and each model.

Table~\ref{tab:slices} gives the scores and intervals behind Table~\ref{tab:groups}, with all tasks pooled.

\begin{table*}[t]
\centering
\small
\setlength{\tabcolsep}{3pt}
\caption{Task completion by scenario and persona group, all tasks pooled, mean over nine models. Change: persona minus original on the same task, with a task-level bootstrap interval. In each of the seven groups, all nine models lose.}
\label{tab:slices}
\begin{tabular}{@{}lrccr@{}}
\toprule
\textbf{Group} & \textbf{Tasks} & \textbf{Original $\to$ persona} & \textbf{Change [95\% CI]} & \textbf{Relative change (\%)} \\
\midrule
All                & 74 & 0.689 $\to$ 0.441 & $-0.248$ [$-0.295$, $-0.202$] & $-36$ \\
Work               & 34 & 0.705 $\to$ 0.431 & $-0.274$ [$-0.352$, $-0.197$] & $-39$ \\
Life               & 40 & 0.675 $\to$ 0.449 & $-0.226$ [$-0.280$, $-0.174$] & $-33$ \\
\midrule
\textsc{expert}    & 74 & 0.689 $\to$ 0.504 & $-0.185$ [$-0.240$, $-0.135$] & $-27$ \\
\textsc{senior}    & 74 & 0.689 $\to$ 0.402 & $-0.287$ [$-0.343$, $-0.232$] & $-42$ \\
\textsc{young}     & 74 & 0.689 $\to$ 0.414 & $-0.275$ [$-0.326$, $-0.225$] & $-40$ \\
\textsc{practical} & 74 & 0.689 $\to$ 0.445 & $-0.243$ [$-0.302$, $-0.187$] & $-35$ \\
\bottomrule
\end{tabular}

\end{table*}

Table~\ref{tab:main} in the main text and Figure~\ref{fig:persona-model} average the per-domain results of Tables~\ref{tab:full-completion} and~\ref{tab:full-satisfaction} over the 74 tasks. Pooled over task types, \textsc{expert} requests reach the highest task completion under all nine models, 0.100 to 0.129 above the lowest group, consistent with the smallest loss for \textsc{expert} in Table~\ref{tab:groups}. Satisfaction is highest for \textsc{expert} under eight of nine models and lowest for \textsc{senior} under seven.

\begin{figure*}[p]
\centering
\includegraphics[width=\textwidth]{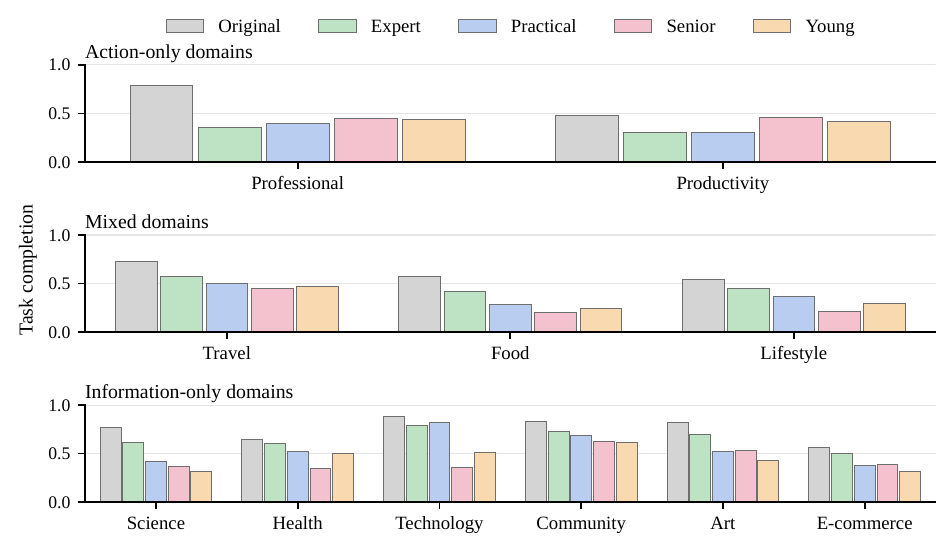}
\caption{Task completion by domain for the original request and each persona group, mean over nine models. The original request is highest in every domain. Table~\ref{tab:full-completion} gives the values for each model.}
\label{fig:completion-control-domain}
\end{figure*}

\begin{figure*}[p]
\centering
\includegraphics[width=\textwidth]{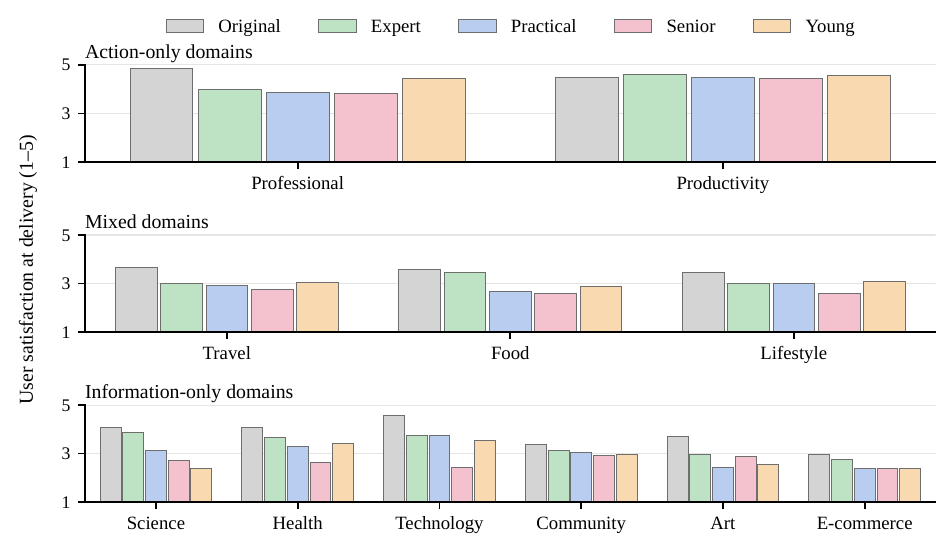}
\caption{Satisfaction at delivery by domain for the original request and each persona group, mean over nine models. The original request is rated by the same model without a profile. Table~\ref{tab:full-satisfaction} gives the values for each model.}
\label{fig:satisfaction-control-domain}
\end{figure*}

Tables~\ref{tab:full-completion} and~\ref{tab:full-satisfaction} include the original request alongside the four persona groups in each domain.
Figures~\ref{fig:completion-control-domain} and~\ref{fig:satisfaction-control-domain} summarize the corresponding domain-level means across the nine backbones, showing the five conditions side by side.
\begin{table*}[p]
\centering
\caption{Task completion by domain, user group, and model. Mixed domains list their information and action tasks. Model columns follow the logos in order of release, as in Table~\ref{tab:main}: MiniMax M3, Kimi K3, Muse Glimmer 30B, Grok 4.6, Qwen 3.8 27B, Gemini 3.8 Flash, GPT-6 Astra, DeepSeek V4.1 Flash, and Claude Opus 5.5. Bold: best model in each row.}
\label{tab:full-completion}
\begingroup
\fontsize{6.5}{7.6}\selectfont
\setlength{\tabcolsep}{1.6pt}
\renewcommand{\arraystretch}{0.92}
\begin{tabular*}{\textwidth}{@{\extracolsep{\fill}}clrrrrrrrrrr@{}}
\toprule
Domain & User &
\multicolumn{1}{c}{\vendoricon{minimax}} &
\multicolumn{1}{c}{\vendoricon{kimi}} &
\multicolumn{1}{c}{\vendoricon{muse}} &
\multicolumn{1}{c}{\vendoricon{grok}} &
\multicolumn{1}{c}{\vendoricon{qwen}} &
\multicolumn{1}{c}{\vendoricon{gemini}} &
\multicolumn{1}{c}{\vendoricon{gpt}} &
\multicolumn{1}{c}{\vendoricon{deepseek}} &
\multicolumn{1}{c}{\vendoricon{claude}} &
Mean \\
\midrule
\multirow{5}{*}{Professional (14)} & \textsc{Original} & 0.857 & \textbf{0.929} & 0.714 & 0.786 & 0.857 & 0.786 & 0.786 & 0.643 & 0.714 & 0.786 \\
  & \textsc{expert} & 0.357 & 0.429 & 0.214 & \textbf{0.571} & 0.357 & 0.286 & 0.357 & 0.214 & 0.429 & 0.357 \\
  & \textsc{senior} & 0.357 & 0.571 & 0.357 & \textbf{0.714} & 0.500 & 0.429 & 0.286 & 0.429 & 0.429 & 0.452 \\
  & \textsc{young} & 0.500 & 0.500 & 0.429 & 0.500 & \textbf{0.571} & 0.429 & 0.286 & 0.214 & 0.500 & 0.437 \\
  & \textsc{practical} & 0.357 & 0.500 & 0.214 & \textbf{0.643} & 0.571 & 0.429 & 0.286 & 0.286 & 0.286 & 0.397 \\
\specialrule{0.3pt}{0.8pt}{0.8pt}
\multirow{5}{*}{Productivity (9)} & \textsc{Original} & 0.444 & \textbf{0.667} & 0.333 & 0.444 & 0.444 & 0.556 & 0.333 & 0.556 & 0.556 & 0.481 \\
  & \textsc{expert} & \textbf{0.333} & \textbf{0.333} & \textbf{0.333} & \textbf{0.333} & 0.222 & \textbf{0.333} & \textbf{0.333} & 0.222 & \textbf{0.333} & 0.309 \\
  & \textsc{senior} & 0.444 & 0.444 & 0.333 & 0.444 & 0.556 & 0.333 & 0.333 & \textbf{0.667} & 0.556 & 0.457 \\
  & \textsc{young} & 0.333 & 0.556 & 0.333 & 0.333 & 0.444 & \textbf{0.625} & 0.333 & 0.333 & 0.444 & 0.415 \\
  & \textsc{practical} & 0.333 & \textbf{0.444} & 0.222 & 0.222 & 0.333 & 0.333 & 0.333 & 0.333 & 0.222 & 0.309 \\
\specialrule{0.3pt}{0.8pt}{0.8pt}
\multirow{5}{*}{Travel (5+7)} & \textsc{Original} & \textbf{0.856} & 0.764 & 0.445 & 0.769 & 0.686 & 0.842 & 0.692 & 0.761 & 0.744 & 0.729 \\
  & \textsc{expert} & 0.547 & 0.620 & 0.419 & 0.467 & \textbf{0.797} & 0.564 & 0.511 & 0.586 & 0.619 & 0.570 \\
  & \textsc{senior} & 0.464 & \textbf{0.544} & 0.311 & 0.525 & 0.450 & 0.461 & 0.375 & 0.469 & 0.481 & 0.453 \\
  & \textsc{young} & 0.356 & \textbf{0.603} & 0.353 & 0.464 & 0.511 & 0.483 & 0.467 & 0.453 & 0.550 & 0.471 \\
  & \textsc{practical} & 0.256 & 0.497 & 0.314 & 0.558 & \textbf{0.622} & 0.575 & 0.514 & 0.569 & 0.617 & 0.502 \\
\specialrule{0.3pt}{0.8pt}{0.8pt}
\multirow{5}{*}{Food (4+4)} & \textsc{Original} & 0.554 & \textbf{0.728} & 0.415 & 0.540 & 0.701 & 0.387 & 0.625 & 0.670 & 0.562 & 0.576 \\
  & \textsc{expert} & 0.210 & \textbf{0.607} & 0.281 & 0.464 & 0.482 & 0.291 & 0.406 & 0.545 & 0.469 & 0.417 \\
  & \textsc{senior} & 0.098 & 0.174 & 0.120 & 0.228 & 0.143 & 0.125 & 0.308 & 0.317 & \textbf{0.321} & 0.204 \\
  & \textsc{young} & 0.219 & 0.339 & 0.165 & 0.245 & \textbf{0.370} & 0.094 & 0.210 & 0.353 & 0.179 & 0.242 \\
  & \textsc{practical} & 0.304 & \textbf{0.402} & 0.165 & 0.339 & 0.308 & 0.238 & 0.277 & 0.321 & 0.245 & 0.289 \\
\specialrule{0.3pt}{0.8pt}{0.8pt}
\multirow{5}{*}{Lifestyle (3+2)} & \textsc{Original} & 0.500 & \textbf{0.645} & 0.515 & 0.555 & 0.620 & \textbf{0.645} & 0.510 & 0.605 & 0.275 & 0.541 \\
  & \textsc{expert} & 0.245 & 0.575 & 0.380 & 0.515 & \textbf{0.605} & 0.485 & 0.325 & 0.470 & 0.485 & 0.454 \\
  & \textsc{senior} & 0.105 & 0.180 & 0.205 & 0.315 & 0.330 & \textbf{0.355} & 0.315 & 0.065 & 0.080 & 0.217 \\
  & \textsc{young} & 0.305 & 0.180 & 0.130 & 0.540 & \textbf{0.550} & 0.380 & 0.270 & 0.170 & 0.170 & 0.299 \\
  & \textsc{practical} & 0.380 & \textbf{0.550} & 0.105 & 0.445 & 0.065 & 0.525 & 0.510 & 0.380 & 0.310 & 0.363 \\
\specialrule{0.3pt}{0.8pt}{0.8pt}
\multirow{5}{*}{Science (6)} & \textsc{Original} & 0.742 & 0.774 & 0.579 & 0.770 & 0.859 & \textbf{0.911} & 0.746 & 0.710 & 0.863 & 0.773 \\
  & \textsc{expert} & \textbf{0.697} & 0.573 & 0.526 & 0.686 & 0.691 & 0.668 & 0.540 & 0.500 & 0.657 & 0.615 \\
  & \textsc{senior} & 0.422 & 0.365 & 0.341 & \textbf{0.453} & 0.304 & 0.379 & 0.349 & 0.290 & 0.429 & 0.370 \\
  & \textsc{young} & 0.258 & 0.296 & 0.278 & 0.272 & \textbf{0.472} & 0.141 & 0.332 & 0.452 & 0.332 & 0.315 \\
  & \textsc{practical} & 0.397 & 0.429 & 0.439 & 0.434 & \textbf{0.490} & 0.314 & 0.417 & 0.434 & 0.452 & 0.423 \\
\specialrule{0.3pt}{0.8pt}{0.8pt}
\multirow{5}{*}{Health (6)} & \textsc{Original} & 0.409 & 0.705 & 0.592 & 0.598 & \textbf{0.752} & 0.635 & 0.700 & 0.663 & 0.714 & 0.641 \\
  & \textsc{expert} & 0.607 & 0.629 & 0.546 & 0.598 & 0.532 & \textbf{0.678} & 0.616 & 0.607 & 0.604 & 0.602 \\
  & \textsc{senior} & 0.234 & 0.300 & 0.318 & 0.452 & 0.421 & 0.186 & 0.366 & 0.382 & \textbf{0.473} & 0.348 \\
  & \textsc{young} & 0.402 & 0.481 & 0.493 & 0.527 & 0.523 & 0.364 & \textbf{0.618} & 0.494 & 0.570 & 0.497 \\
  & \textsc{practical} & 0.495 & 0.513 & 0.402 & 0.564 & 0.556 & 0.447 & \textbf{0.649} & 0.532 & 0.552 & 0.523 \\
\specialrule{0.3pt}{0.8pt}{0.8pt}
\multirow{5}{*}{Tech (4)} & \textsc{Original} & 0.829 & 0.778 & 0.864 & 0.829 & 0.950 & 0.879 & \textbf{0.964} & 0.914 & 0.914 & 0.880 \\
  & \textsc{expert} & 0.629 & 0.714 & 0.764 & 0.829 & 0.900 & 0.629 & \textbf{0.914} & 0.864 & 0.850 & 0.788 \\
  & \textsc{senior} & 0.157 & 0.357 & 0.357 & 0.272 & 0.286 & 0.407 & 0.407 & \textbf{0.557} & 0.443 & 0.360 \\
  & \textsc{young} & 0.186 & 0.457 & 0.593 & 0.607 & 0.557 & 0.471 & \textbf{0.643} & 0.493 & 0.593 & 0.511 \\
  & \textsc{practical} & 0.729 & 0.729 & 0.664 & 0.829 & 0.850 & 0.829 & 0.914 & \textbf{0.950} & 0.864 & 0.817 \\
\specialrule{0.3pt}{0.8pt}{0.8pt}
\multirow{5}{*}{Community (4)} & \textsc{Original} & \textbf{0.902} & 0.812 & 0.768 & \textbf{0.902} & 0.839 & \textbf{0.902} & 0.777 & 0.777 & 0.839 & 0.835 \\
  & \textsc{expert} & 0.518 & 0.803 & 0.670 & 0.643 & 0.768 & \textbf{0.866} & 0.768 & 0.830 & 0.705 & 0.730 \\
  & \textsc{senior} & 0.572 & \textbf{0.803} & 0.393 & 0.553 & 0.714 & 0.536 & 0.741 & 0.714 & 0.643 & 0.630 \\
  & \textsc{young} & 0.473 & 0.607 & 0.678 & \textbf{0.741} & \textbf{0.741} & 0.607 & 0.643 & 0.384 & 0.705 & 0.620 \\
  & \textsc{practical} & 0.607 & \textbf{0.803} & 0.589 & 0.768 & 0.741 & 0.616 & 0.705 & 0.705 & 0.643 & 0.686 \\
\specialrule{0.3pt}{0.8pt}{0.8pt}
\multirow{5}{*}{Art (3)} & \textsc{Original} & 0.657 & 0.802 & 0.488 & 0.802 & 0.952 & \textbf{1.000} & 0.917 & \textbf{1.000} & 0.802 & 0.825 \\
  & \textsc{expert} & 0.560 & 0.655 & 0.660 & 0.750 & 0.738 & 0.379 & \textbf{0.917} & 0.738 & 0.869 & 0.696 \\
  & \textsc{senior} & 0.510 & 0.510 & 0.362 & 0.490 & \textbf{0.671} & 0.395 & 0.607 & 0.624 & 0.655 & 0.536 \\
  & \textsc{young} & 0.312 & 0.293 & 0.476 & 0.540 & 0.493 & 0.245 & \textbf{0.671} & 0.512 & 0.343 & 0.432 \\
  & \textsc{practical} & 0.493 & 0.540 & 0.426 & 0.426 & 0.624 & 0.426 & 0.540 & 0.560 & \textbf{0.690} & 0.525 \\
\specialrule{0.3pt}{0.8pt}{0.8pt}
\multirow{5}{*}{E-commerce (3)} & \textsc{Original} & 0.588 & \textbf{0.658} & 0.438 & 0.588 & 0.438 & 0.594 & 0.575 & 0.655 & 0.496 & 0.559 \\
  & \textsc{expert} & 0.505 & 0.552 & 0.413 & 0.480 & 0.438 & 0.438 & \textbf{0.630} & 0.496 & 0.575 & 0.503 \\
  & \textsc{senior} & 0.162 & 0.324 & 0.438 & 0.455 & 0.457 & 0.324 & \textbf{0.505} & 0.455 & 0.346 & 0.385 \\
  & \textsc{young} & 0.390 & 0.276 & \textbf{0.438} & 0.242 & 0.089 & 0.210 & 0.346 & 0.413 & \textbf{0.438} & 0.316 \\
  & \textsc{practical} & 0.343 & 0.114 & 0.480 & \textbf{0.605} & 0.438 & 0.305 & 0.467 & 0.346 & 0.335 & 0.381 \\
\bottomrule
\end{tabular*}
\par\endgroup

\end{table*}

\begin{table*}[p]
\centering
\caption{Satisfaction at delivery by domain, user group, and model. The original request is rated by the same model without a profile. Model columns match Table~\ref{tab:full-completion}. Bold: best model in each row.}
\label{tab:full-satisfaction}
\begingroup
\fontsize{6.5}{7.6}\selectfont
\setlength{\tabcolsep}{1.6pt}
\renewcommand{\arraystretch}{0.92}
\begin{tabular*}{\textwidth}{@{\extracolsep{\fill}}clrrrrrrrrrr@{}}
\toprule
Domain & User &
\multicolumn{1}{c}{\vendoricon{minimax}} &
\multicolumn{1}{c}{\vendoricon{kimi}} &
\multicolumn{1}{c}{\vendoricon{muse}} &
\multicolumn{1}{c}{\vendoricon{grok}} &
\multicolumn{1}{c}{\vendoricon{qwen}} &
\multicolumn{1}{c}{\vendoricon{gemini}} &
\multicolumn{1}{c}{\vendoricon{gpt}} &
\multicolumn{1}{c}{\vendoricon{deepseek}} &
\multicolumn{1}{c}{\vendoricon{claude}} &
Mean \\
\midrule
\multirow{5}{*}{Professional (14)} & \textsc{Original} & 4.79 & \textbf{5.00} & 4.50 & \textbf{5.00} & \textbf{5.00} & \textbf{5.00} & 4.79 & 4.71 & 4.86 & 4.85 \\
  & \textsc{expert} & 3.57 & 3.93 & 3.86 & \textbf{4.50} & 4.07 & 3.57 & 4.21 & 4.14 & 4.07 & 3.99 \\
  & \textsc{senior} & 3.71 & 4.07 & 3.21 & 3.93 & 4.21 & 3.43 & 3.29 & 4.00 & \textbf{4.43} & 3.81 \\
  & \textsc{young} & 4.21 & \textbf{4.93} & 4.14 & 4.71 & 4.57 & 4.21 & 4.36 & 4.57 & 4.21 & 4.44 \\
  & \textsc{practical} & 4.14 & 4.14 & 3.00 & 3.93 & 3.43 & 3.64 & 4.00 & \textbf{4.29} & 4.14 & 3.86 \\
\specialrule{0.3pt}{0.8pt}{0.8pt}
\multirow{5}{*}{Productivity (9)} & \textsc{Original} & 4.67 & 4.67 & 4.78 & 4.56 & 4.11 & \textbf{5.00} & 4.33 & 4.00 & 4.33 & 4.49 \\
  & \textsc{expert} & 4.11 & \textbf{5.00} & \textbf{5.00} & \textbf{5.00} & 3.78 & \textbf{5.00} & 4.33 & 4.33 & 4.78 & 4.59 \\
  & \textsc{senior} & \textbf{5.00} & 4.22 & 4.89 & 4.22 & 4.89 & \textbf{5.00} & 3.56 & 4.22 & 4.11 & 4.46 \\
  & \textsc{young} & 4.33 & 4.67 & 4.89 & 4.44 & 4.11 & \textbf{5.00} & 4.78 & 4.56 & 4.22 & 4.56 \\
  & \textsc{practical} & 4.00 & 4.67 & 4.89 & 4.89 & 4.00 & \textbf{5.00} & 4.67 & 3.89 & 4.44 & 4.49 \\
\specialrule{0.3pt}{0.8pt}{0.8pt}
\multirow{5}{*}{Travel (5+7)} & \textsc{Original} & 3.83 & 3.75 & 2.58 & \textbf{4.17} & 3.92 & \textbf{4.17} & 3.17 & 3.83 & 3.42 & 3.65 \\
  & \textsc{expert} & 2.67 & 3.17 & 2.58 & 3.42 & \textbf{3.50} & 3.00 & 2.67 & 3.00 & 3.08 & 3.01 \\
  & \textsc{senior} & \textbf{3.08} & 2.83 & 2.58 & 2.75 & 2.83 & 2.42 & 2.92 & 2.75 & 2.50 & 2.74 \\
  & \textsc{young} & 2.42 & 3.25 & 2.58 & 3.25 & \textbf{3.33} & 3.00 & 2.92 & \textbf{3.33} & 3.17 & 3.03 \\
  & \textsc{practical} & 2.42 & 2.75 & 2.25 & 3.17 & \textbf{3.42} & 3.00 & 2.92 & 3.33 & 3.17 & 2.94 \\
\specialrule{0.3pt}{0.8pt}{0.8pt}
\multirow{5}{*}{Food (4+4)} & \textsc{Original} & 4.00 & \textbf{4.12} & 3.25 & 3.38 & 3.50 & 3.17 & 3.62 & 3.62 & 3.50 & 3.57 \\
  & \textsc{expert} & 3.00 & \textbf{3.88} & 3.25 & 3.25 & \textbf{3.88} & 3.29 & 3.38 & \textbf{3.88} & 3.38 & 3.46 \\
  & \textsc{senior} & 2.38 & 2.75 & 2.12 & 2.62 & 2.75 & 2.50 & 2.38 & \textbf{3.50} & 2.50 & 2.61 \\
  & \textsc{young} & 2.38 & 3.12 & 2.62 & 3.00 & \textbf{3.50} & 2.75 & 2.25 & 3.25 & 2.88 & 2.86 \\
  & \textsc{practical} & 2.50 & 2.62 & 2.88 & 2.88 & 2.50 & 2.50 & 2.25 & \textbf{3.12} & 2.88 & 2.68 \\
\specialrule{0.3pt}{0.8pt}{0.8pt}
\multirow{5}{*}{Lifestyle (3+2)} & \textsc{Original} & \textbf{3.80} & 3.40 & 3.40 & 3.40 & 2.80 & 3.60 & \textbf{3.80} & 3.60 & 3.20 & 3.44 \\
  & \textsc{expert} & 2.20 & \textbf{4.00} & 2.20 & 3.00 & 3.00 & 2.60 & 3.40 & 3.60 & 3.20 & 3.02 \\
  & \textsc{senior} & 2.60 & 3.20 & 2.20 & 2.80 & \textbf{3.40} & 1.80 & 3.00 & 2.20 & 2.00 & 2.58 \\
  & \textsc{young} & 3.20 & 3.20 & 2.60 & \textbf{3.60} & \textbf{3.60} & 3.40 & 3.20 & 2.40 & 2.60 & 3.09 \\
  & \textsc{practical} & \textbf{3.80} & 3.60 & 2.00 & 3.00 & 1.80 & \textbf{3.80} & 3.40 & 2.60 & 3.20 & 3.02 \\
\specialrule{0.3pt}{0.8pt}{0.8pt}
\multirow{5}{*}{Science (6)} & \textsc{Original} & 3.50 & 4.67 & 3.17 & 4.00 & 4.50 & \textbf{4.83} & 4.00 & 4.00 & 4.17 & 4.09 \\
  & \textsc{expert} & 3.17 & \textbf{4.50} & 3.33 & 4.33 & 4.17 & 3.50 & 4.17 & 3.33 & 4.17 & 3.85 \\
  & \textsc{senior} & 2.67 & 2.50 & 3.00 & \textbf{3.17} & 2.50 & 2.33 & 2.83 & 2.67 & 2.67 & 2.70 \\
  & \textsc{young} & \textbf{2.67} & 2.50 & 2.17 & 2.33 & \textbf{2.67} & 2.17 & 2.17 & 2.50 & 2.50 & 2.41 \\
  & \textsc{practical} & 2.33 & 3.00 & 2.67 & 3.50 & 3.33 & 3.00 & \textbf{3.83} & 3.00 & 3.50 & 3.13 \\
\specialrule{0.3pt}{0.8pt}{0.8pt}
\multirow{5}{*}{Health (6)} & \textsc{Original} & 3.83 & 4.33 & 4.00 & 4.17 & 4.17 & \textbf{4.67} & 4.17 & 3.67 & 3.67 & 4.07 \\
  & \textsc{expert} & 2.83 & \textbf{4.50} & 2.67 & 3.67 & \textbf{4.50} & 3.67 & 3.67 & 3.67 & 3.67 & 3.65 \\
  & \textsc{senior} & 2.00 & 2.83 & 2.67 & 2.67 & \textbf{3.17} & 2.33 & 2.83 & 2.33 & 2.83 & 2.63 \\
  & \textsc{young} & 2.67 & 3.50 & 2.83 & 3.50 & \textbf{3.83} & 3.33 & 3.67 & \textbf{3.83} & 3.50 & 3.41 \\
  & \textsc{practical} & 2.83 & 3.83 & 2.67 & 3.17 & \textbf{4.00} & 3.17 & 3.50 & 3.50 & 3.17 & 3.31 \\
\specialrule{0.3pt}{0.8pt}{0.8pt}
\multirow{5}{*}{Tech (4)} & \textsc{Original} & 4.25 & 4.75 & 4.50 & 4.50 & \textbf{5.00} & 4.25 & 4.25 & \textbf{5.00} & 4.50 & 4.56 \\
  & \textsc{expert} & 3.00 & 3.50 & 4.00 & \textbf{4.75} & 4.25 & 2.25 & \textbf{4.75} & 3.50 & 3.75 & 3.75 \\
  & \textsc{senior} & 1.75 & 2.50 & 2.25 & 2.50 & \textbf{2.75} & 2.25 & 2.50 & \textbf{2.75} & \textbf{2.75} & 2.44 \\
  & \textsc{young} & 2.50 & 3.25 & 3.50 & \textbf{4.50} & 4.25 & 2.75 & 3.25 & 3.25 & \textbf{4.50} & 3.53 \\
  & \textsc{practical} & 3.00 & 4.00 & 2.75 & \textbf{4.50} & 4.00 & 2.75 & \textbf{4.50} & 4.00 & 4.25 & 3.75 \\
\specialrule{0.3pt}{0.8pt}{0.8pt}
\multirow{5}{*}{Community (4)} & \textsc{Original} & 3.50 & 3.25 & 2.75 & \textbf{3.75} & 3.50 & 3.50 & 3.50 & 3.50 & 3.00 & 3.36 \\
  & \textsc{expert} & 3.00 & 3.25 & 2.25 & 3.25 & \textbf{3.75} & 2.75 & \textbf{3.75} & 3.00 & 3.00 & 3.11 \\
  & \textsc{senior} & 3.00 & 3.25 & 2.50 & 2.50 & 3.00 & 2.50 & \textbf{3.50} & 3.00 & 3.00 & 2.92 \\
  & \textsc{young} & 2.75 & 2.75 & 2.25 & \textbf{3.50} & 3.00 & 3.00 & 3.25 & 3.00 & 3.25 & 2.97 \\
  & \textsc{practical} & 3.25 & 3.50 & 2.50 & 3.00 & 3.00 & 3.00 & \textbf{3.75} & 2.75 & 2.75 & 3.06 \\
\specialrule{0.3pt}{0.8pt}{0.8pt}
\multirow{5}{*}{Art (3)} & \textsc{Original} & 2.67 & 3.67 & 2.33 & 4.00 & \textbf{4.67} & \textbf{4.67} & \textbf{4.67} & 3.67 & 3.00 & 3.70 \\
  & \textsc{expert} & 2.67 & 2.33 & 3.00 & 2.67 & 2.67 & 2.67 & \textbf{3.67} & 3.33 & \textbf{3.67} & 2.96 \\
  & \textsc{senior} & 2.33 & 3.00 & 2.33 & 2.67 & 3.33 & 2.67 & 3.33 & \textbf{3.67} & 2.67 & 2.89 \\
  & \textsc{young} & 2.33 & 2.33 & 2.00 & 3.00 & 3.00 & 2.00 & \textbf{3.33} & 2.67 & 2.33 & 2.56 \\
  & \textsc{practical} & 2.00 & 2.67 & 2.00 & 2.33 & 2.33 & 2.67 & 2.67 & \textbf{3.00} & 2.33 & 2.44 \\
\specialrule{0.3pt}{0.8pt}{0.8pt}
\multirow{5}{*}{E-commerce (3)} & \textsc{Original} & 2.67 & 3.33 & 2.33 & 3.00 & 2.33 & 2.67 & \textbf{3.67} & 3.33 & 3.33 & 2.96 \\
  & \textsc{expert} & 2.00 & 3.33 & 2.67 & 3.00 & 2.67 & 2.00 & \textbf{3.67} & 3.00 & 2.67 & 2.78 \\
  & \textsc{senior} & 2.00 & 2.33 & 1.67 & 2.67 & 2.33 & 2.00 & 3.00 & 2.33 & \textbf{3.33} & 2.41 \\
  & \textsc{young} & 2.33 & 2.00 & 2.67 & 1.67 & 2.33 & 2.00 & \textbf{3.00} & 2.33 & \textbf{3.00} & 2.37 \\
  & \textsc{practical} & 1.67 & 2.00 & 2.00 & 3.00 & 2.67 & 2.00 & \textbf{3.33} & 2.33 & 2.67 & 2.41 \\
\bottomrule
\end{tabular*}
\par\endgroup

\end{table*}

\subsection{Task Completion vs. User Satisfaction}
\label{app:res-report}
Satisfaction (1--5) is rated by the persona agent, which is given the full dialogue of its own rollout and answers in the first person whether it is satisfied and whether it believes the task was done. The external judge is the P5 score of Appendix~\ref{app:eval}. Table~\ref{tab:satisfaction} reports persona cells only. Under a persona the external judge barely separates success from failure: its AUC on action tasks is 0.56 (0.65 on original requests) and its correlation with the information score is 0.33 (0.56 on original requests). The user's own rating reaches an AUC of 0.62 at delivery and 0.76 after verification, and a correlation of 0.59. Satisfaction on failed action cells falls from 3.55 at delivery to 2.39 after verification. Information answers can be verified only by reading the cited pages, so we do not collect after-verification ratings for them.

\begin{table*}[t]
\centering
\caption{How well each rating tracks task completion, on persona cells. AUC measures separation of passed from failed action cells (0.5 is chance). Information correlation is the correlation between the rating and the information-task score. At delivery the user sees the final reply and the open tabs, and when verified the user also sees the resulting site state. Verified ratings cover only a subset of the tasks, namely those that change the site. Misreport is the share of loss cells whose final reply is inconsistent with the outcome.}
\label{tab:satisfaction}
\resizebox{\textwidth}{!}{%
\begin{tabular}{@{}l ccc cc cc cc@{}}
\toprule
\multicolumn{1}{c}{\multirow[c]{2}{*}[-0.6ex]{\textbf{Model}}} & \multicolumn{3}{c}{\textbf{Action AUC}} & \multicolumn{2}{c}{\textbf{Failed, judged done (\%)}} & \multicolumn{2}{c}{\textbf{Information correlation}} & \multicolumn{2}{c}{\textbf{Misreport (\%)}} \\
\cmidrule(lr){2-4}\cmidrule(lr){5-6}\cmidrule(lr){7-8}\cmidrule(lr){9-10}
 & External judge & Delivery & Verified & Delivery & Verified & External judge & Delivery & Information & Action \\
\midrule
\vendoricon{minimax}MiniMax M3          & 0.56 & 0.66 & 0.76 & 54 & 33 & 0.28 & 0.51 & 58 & 38 \\
\vendoricon{kimi}Kimi K3             & 0.58 & 0.62 & 0.78 & 62 & 28 & 0.29 & 0.54 & 49 & 22 \\
\vendoricon{muse}Muse Glimmer 30B    & 0.66 & 0.73 & 0.81 & 42 & 25 & 0.46 & 0.63 & 60 & 33 \\
\vendoricon{grok}Grok 4.6            & 0.49 & 0.55 & 0.69 & 64 & 38 & 0.34 & 0.62 & 36 & 17 \\
\vendoricon{qwen}Qwen 3.8 27B        & 0.56 & 0.63 & 0.75 & 58 & 30 & 0.27 & 0.57 & 43 & 32 \\
\vendoricon{gemini}Gemini 3.8 Flash    & 0.61 & 0.59 & 0.82 & 59 & 21 & 0.22 & 0.51 & 60 & 28 \\
\vendoricon{gpt}GPT-6 Astra         & 0.55 & 0.64 & 0.78 & 48 & 16 & 0.23 & 0.50 & 12 & 14 \\
\vendoricon{deepseek}DeepSeek V4.1 Flash & 0.50 & 0.52 & 0.65 & 68 & 24 & 0.33 & 0.70 & 54 & 39 \\
\vendoricon{claude}Claude Opus 5.5     & 0.58 & 0.63 & 0.75 & 58 & 31 & 0.26 & 0.56 & 24 & 6  \\
\midrule
\textbf{Pooled}     & 0.56 & 0.62 & 0.76 & 56 & 27 & 0.33 & 0.59 & 45 & 26 \\
\bottomrule
\end{tabular}}

\end{table*}

\begin{wraptable}{R}{0.48\textwidth}
\centering
\small
\setlength{\tabcolsep}{1.25pt}
\caption{Satisfaction rating from 1 to 5 and share of cells the user judges done, for the 343 action loss cells. Verified ratings are given after the user checks the site and cover only the tasks that change a site.}
\label{tab:stops}
\begin{tabular*}{\linewidth}{@{\extracolsep{\fill}}l c cc cc@{}}
\toprule
\multicolumn{1}{c}{\multirow[c]{2}{*}[-0.6ex]{\textbf{Group}}} & \multicolumn{1}{c}{\multirow[c]{2}{*}[-0.6ex]{\textbf{Cells}}} & \multicolumn{2}{c}{\textbf{At delivery}} & \multicolumn{2}{c}{\textbf{Verified}} \\
\cmidrule(lr){3-4}\cmidrule(l){5-6}
 & & Rating & Done (\%) & Rating & Done (\%) \\
\midrule
Stopped correctly        & 138 & 3.69 & 62 & 1.86 & 7 \\
Other failures           & 205 & 3.25 & 46 & 2.42 & 27 \\
\quad reply misreports   & 75  & 3.47 & 55 & 2.63 & 33 \\
\quad reply is accurate  & 130 & 3.13 & 41 & 2.31 & 24 \\
\bottomrule
\end{tabular*}
\end{wraptable}
Table~\ref{tab:stops} separates the action loss cells in which the agent stopped correctly from the other failures. A stop counts as correct when the final site state shows that nothing was submitted and the labeler finds that the user asked for a draft, a review, or to finish themselves. In the former, the user asked only for a draft or asked to review first, and 62\% judge the task done at delivery, which from their perspective it is; after verification the site shows no change and satisfaction falls to 1.86, so the after-verification rating counts a correct stop as a failure.

The verified rating scores these stops against the original goal. At delivery the user sees the draft and a reply stating that nothing was sent, which is what it asked for, and 62\% judge the task done. After verification it is also shown that the site did not change, while it still holds the original instruction $x_t$, which asks for the submission, and only 7\% judge the task done. The verified rating therefore does not separate a stop the user asked for from a failure to act. We treat the rating at delivery as the measure of whether the user was served, and the verified rating as a check against the original goal.

\textcolor{black}{The main-experiment ratings and labels in this appendix come from GPT-5.6-terra, which also plays the persona agent and the external judge. Appendix~\ref{app:human} checks requirement met, raised, and engaged and persona fidelity. Appendix~\ref{app:sat-check} additionally compares the persona agent's own satisfaction at delivery and after verification, and the external judge's P5 score, with ratings from the same two human annotators, P1 and P2.}

\begin{table}[t]
\centering
\small
\setlength{\tabcolsep}{1.25pt}
\caption{Satisfaction from 1 to 5 on each group's own successful and failed cells, pooled over nine models and all tasks. Yes and No mark whether the cell is fully successful. After-verification ratings cover only the tasks that change a site.}
\label{tab:sat-outcome}
\begin{tabular*}{0.7\linewidth}{@{\extracolsep{\fill}}l c cc cc@{}}
\toprule
\multicolumn{1}{c}{\multirow[c]{2}{*}[-0.6ex]{\textbf{Group}}} & \multicolumn{1}{c}{\multirow[c]{2}{*}[-0.6ex]{\tablehead{\textbf{Fully}\\\textbf{successful (\%)}}}} & \multicolumn{2}{c}{\textbf{At delivery}} & \multicolumn{2}{c}{\textbf{Verified}} \\
\cmidrule(lr){3-4}\cmidrule(l){5-6}
 & & Yes & No & Yes & No \\
\midrule
Original            & 47.4 & 4.68 & 3.46 & 4.38 & 2.65 \\
\midrule
\persona{expert}    & 27.4 & 4.03 & 3.43 & 3.96 & 2.23 \\
\persona{senior}    & 22.5 & 4.11 & 2.81 & 3.80 & 2.40 \\
\persona{young}     & 22.9 & 4.49 & 3.11 & 4.25 & 2.57 \\
\persona{practical} & 24.4 & 4.00 & 3.12 & 3.70 & 2.38 \\
\bottomrule
\end{tabular*}
\end{table}
Table~\ref{tab:sat-outcome} splits each group's cells by their own outcome. A cell is fully successful when the state check passes on an action task and every requirement is met on an information task. Every rating is given by the persona agent that made the request, and the original request is rated by the same model without a profile. Persona requests are fully successful far less often, in 22.5\% to 27.4\% of cells against 47.4\% for the original request. On successful cells, satisfaction at delivery is 4.00 to 4.49 under a persona against 4.68 for the original request, and on failed cells it is 2.81 to 3.43 against 3.46. After verification, failed cells fall to between 2.23 and 2.65 in every group. Among fully successful cells, the share rated 5 at delivery is 84.1\% for the original request, 57.1\% for \textsc{expert}, 54.9\% for \textsc{practical}, 64.0\% for \textsc{senior}, and 76.3\% for \textsc{young}. Over all successful persona cells it is 62.7\%, and 25.9\% are rated 3 or lower, against 9.5\% of successful original-request cells.

\subsection{Where the Drop Comes From}
\label{app:res-loss}
Table~\ref{tab:labels} gives the labels behind the primary causes of Figure~\ref{fig:loss-causes}. Unlike the primary causes, a cell can carry several labels, and the table splits them by task type.

\begin{table}[t]
\centering
\small
\setlength{\tabcolsep}{3pt}
\caption{Labels of the 887 information and 343 action loss cells. A cell can carry several labels. The final step is to submit, book, or send. Labels marked $^{a}$ come from the per-requirement scores, those marked $^{b}$ from the final site state, and the rest from GPT-5.6-terra.}
\label{tab:labels}
\begin{tabular}{@{}>{\raggedright\arraybackslash}m{0.60\linewidth}rr@{}}
\toprule
\textbf{Label (\% of loss cells)} & \textbf{Information} & \textbf{Action} \\
\midrule
Missed a raised requirement$^{a}$ & 76.7 & 24.2 \\
Missed a requirement not raised$^{a}$ & 45.7 & 1.7 \\
\quad only requirements not raised missed$^{a}$ & 23.3 & 0.9 \\
Final step not executed$^{b}$ & -- & 74.9 \\
\quad of which a correct stop & -- & 40.2 \\
User asked for less or for staged execution & 14.7 & 59.2 \\
Persona agent gave wrong information & 1.5 & 4.1 \\
Key question left unresolved & 4.4 & 15.7 \\
Final reply inconsistent with the outcome & 45.1 & 25.9 \\
\midrule
Complied with the current request (yes\,/\,partly\,/\,no) & 38/61/1 & 56/35/9 \\
\bottomrule
\end{tabular}
\end{table}
The primary causes of Figure~\ref{fig:loss-causes} are not labeled separately. They follow by rule from three sources: the per-requirement raised, engaged, and met labels; the final site state, which marks a stop; and the GPT-5.6-terra label for whether the user asked for a draft, a review, or to finish the action themselves. Their reliability is that of these labels. Agreement with the other annotators is higher for met than for raised and engaged (Table~\ref{tab:human}), and raised and engaged decide the split between losses on raised requirements (59.9\%) and losses only on requirements the user never raised (17.1\%). The primary cause itself is not re-annotated.

The shares count loss cells, not points of task completion. A cell that loses 0.1 and a cell that loses 1.0 count the same, so 59.9\% is the share of loss cells whose primary cause is a raised requirement that the agent engaged with and still missed, not the share of the drop in task completion that this cause explains.

Task completion is scored against the original instruction, while a persona request may omit a requirement, add a constraint, or ask for a draft first (\S\ref{sec:rewrite}). The drop in Table~\ref{tab:groups} is therefore a drop in task completion on the original specification, and it combines failures of execution with changes in what the user asked for and with errors of the persona agent. The primary causes separate part of this mixture. In 11.2\% of loss cells the agent stopped because the user asked it to, and in 17.1\% the cell lost points only on requirements the user never raised, so 28.3\% of loss cells reflect a changed or unstated request rather than a failure on what the user asked for. The persona agent gives wrong information in 1.5\% of information and 4.1\% of action loss cells (Table~\ref{tab:labels}), and excluding the 11.1\% of cells in which it contradicts its profile changes the mean $\Delta$ from $-0.248$ to $-0.249$ (Appendix~\ref{app:fidelity}).

\begin{figure}[t]
\centering
\includegraphics[width=0.6\linewidth]{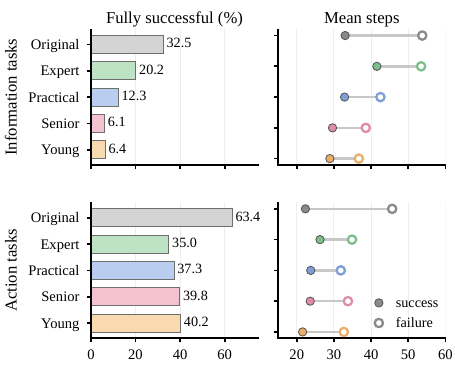}
\caption{Share of fully successful cells and mean steps of successful and failed cells, by user group, pooled over nine models. Only 21 information cells are fully successful for \textsc{senior} and 22 for \textsc{young}.}
\label{fig:steps}
\end{figure}
(i)~Correct stop (Qwen 3.8 27B, professional, \textsc{expert}): the user asks only for a draft reply; the agent provides the draft and does not send it, and the checker scores a failure. (ii)~Only requirements not raised (Kimi K3, health, \textsc{senior}): the user replaces three named businesses with a request for a gym nearby; the agent finds a different one, which satisfies the user and fails every rubric requirement. (iii)~Changed request (Qwen 3.8 27B, food, \textsc{senior}): the user adds a dietary preference and a preference for a nearby place and says they will enter the contact details themselves; the agent does not submit, no reservation is created, and the neighborhood named in the original instruction is lost. (iv)~Unresolved question and misreport (Gemini 3.8 Flash, travel, \textsc{senior}): the user does not state the year, the agent books for 2025, the passenger form fails, and the final reply states that checkout has been arranged.

Figure~\ref{fig:steps} relates outcome to episode length. The outcome here is whether a cell is fully successful: an action cell passes the state check, and an information cell meets every requirement. The main text scores information cells by the share of requirements met, so the original request is fully successful in 47.4\% of cells against a mean task completion of 0.689. Pooled over task types, failed persona cells take 35.2 to 45.4 steps against 51.0 for failed original-request cells; the gap holds for every group on action tasks (31.8 to 34.9 against 45.7) and on information tasks for every group except \textsc{expert} (53.5 against 53.8). Failed persona cells thus end early, which fits the agent stopping before the final step (Table~\ref{tab:labels}).

\subsection{When Agents Proactively Question}
\label{app:res-ask}
\S\ref{sec:res-form} showed that users leave some requirements unsaid, and an agent can recover such requirements by asking. Although omissions concentrate on information tasks, only GPT-6 Astra and Claude Opus 5.5 ask a clarification question in more than 5\% of these cells, at 23.0\% and 11.8\% as Table~\ref{tab:behavior-overview} shows. Of the 609 questions asked by all models, 24\% target a requirement the user omitted, 53\% concern one already stated, and 23\% are unrelated to any requirement.

The persona agent limits what a question can recover. 78\% of its answers add details that are absent from the task and 9\% conflict with it, and cells with a conflicting answer score 0.14 against 0.29 for the others. Because agents ask more often on tasks that are harder to begin with, \S\ref{sec:res-force} tests the effect of asking by intervention. Table~\ref{tab:clarification} reports the comparison between cells with and without a question.

\begin{table*}[p]
\centering
\small
\setlength{\tabcolsep}{0pt}
\caption{Clarification and outcome reporting by model, in the order of Table~\ref{tab:main}. Asked a question counts persona cells, misreport counts loss cells, and judged done counts failed action cells. $^{\dagger}$Pooled over cells.}
\label{tab:behavior-overview}
\begin{tabular*}{\textwidth}{@{\extracolsep{\fill}}l cc cc c@{}}
\toprule
\multicolumn{1}{c}{\multirow[c]{2}{*}[-0.6ex]{\textbf{Model}}} & \multicolumn{2}{c}{\textbf{Asked a question (\%)}} & \multicolumn{2}{c}{\textbf{Misreport (\%)}} & \multirow[c]{2}{*}[-0.6ex]{\tablehead{\textbf{Judged done}\\(\%)}} \\
\cmidrule(lr){2-3}\cmidrule(lr){4-5}
 & Information & Action & Information & Action & \\
\midrule
\vendoricon{minimax}MiniMax M3           & 4.6  & 40.3 & 58 & 38 & 54 \\
\vendoricon{kimi}Kimi K3                 & 4.6  & 22.9 & 49 & 22 & 62 \\
\vendoricon{muse}Muse Glimmer 30B        & 3.3  & 16.7 & 60 & 33 & 42 \\
\vendoricon{grok}Grok 4.6                & 1.3  & 2.8  & 36 & 17 & 64 \\
\vendoricon{qwen}Qwen 3.8 27B            & 3.3  & 27.1 & 43 & 32 & 58 \\
\vendoricon{gemini}Gemini 3.8 Flash      & 0.0  & 6.4  & 60 & 28 & 59 \\
\vendoricon{gpt}GPT-6 Astra              & 23.0 & 53.5 & 12 & 14 & 48 \\
\vendoricon{deepseek}DeepSeek V4.1 Flash & 2.0  & 31.2 & 54 & 39 & 68 \\
\vendoricon{claude}Claude Opus 5.5       & 11.8 & 49.3 & 24 & 6  & 58 \\
\midrule
\textbf{Mean}                          & --   & --   & 45$^{\dagger}$ & 26$^{\dagger}$ & 56$^{\dagger}$ \\
\bottomrule
\end{tabular*}

\end{table*}

Table~\ref{tab:behavior-overview} summarizes clarification and outcome reporting for each model, and Table~\ref{tab:clarification} gives the details of clarification. On action tasks, agents ask in 78\% of the cells in which the user asks to review first, mostly to confirm, in 16\% of those in which the user asks only for help, and in 28\% of those in which the user asks the agent to act. Agents ask more often on tasks whose original request also fails, 31\% against 26\% on action tasks and 7.5\% against 2.9\% on information tasks. When a question does target an omitted requirement, that requirement is eventually met in 10 of 33 cases on information tasks, against 61\% for the same requirements in original-request cells, and in 14 of 20 cases on action tasks. The within-task difference between cells with and without a question is positive for four models and negative for five. In one typical failure, the agent asks which year is meant, all four personas answer 2025 or 2026 while the site contains data for 2024, and all four cells fail. The original request, in which no question is asked, passes.

\begin{table*}[p]
\centering
\caption{Clarification behavior. Ask is the share of cells with at least one question. On target counts questions that target a requirement the user omitted. Invented counts persona-agent answers that add details absent from the task, and Conflict counts answers whose details conflict with the task. Difference is, within a task, mean task completion of persona cells in which the agent asked minus that of cells in which it did not, with the number of tasks containing both kinds of cell.}
\label{tab:clarification}
\resizebox{\textwidth}{!}{%
\begin{tabular}{@{}l cc cc r r cc c@{}}
\toprule
\multicolumn{1}{c}{\multirow[c]{2}{*}[-0.6ex]{\textbf{Model}}} & \multicolumn{2}{c}{\textbf{Ask, information (\%)}} & \multicolumn{2}{c}{\textbf{Ask, action (\%)}} & \multicolumn{1}{c}{\multirow[c]{2}{*}[-0.6ex]{\textbf{Questions}}} & \multicolumn{1}{c}{\multirow[c]{2}{*}[-0.6ex]{\tablehead{\textbf{On}\\\textbf{target (\%)}}}} & \multicolumn{2}{c}{\textbf{Answers (\%)}} & \multirow[c]{2}{*}[-0.6ex]{\tablehead{\textbf{Asked $-$ not asked}\\Difference (tasks)}} \\
\cmidrule(lr){2-3}\cmidrule(lr){4-5}\cmidrule(lr){8-9}
 & Persona & Original & Persona & Original & & & Invented & Conflict & \\
\midrule
\vendoricon{minimax}MiniMax M3          & 4.6  & 5.3  & 40.3 & 25.0 & 96  & 21 & 73 & 3  & $+0.103$ (26) \\
\vendoricon{kimi}Kimi K3             & 4.6  & 2.6  & 22.9 & 8.3  & 55  & 36 & 73 & 11 & $+0.164$ (16) \\
\vendoricon{muse}Muse Glimmer 30B    & 3.3  & 5.3  & 16.7 & 8.3  & 31  & 10 & 97 & 0  & $-0.185$ (9)  \\
\vendoricon{grok}Grok 4.6            & 1.3  & 0.0  & 2.8  & 0.0  & 9   & 33 & 89 & 22 & $-0.500$ (5)  \\
\vendoricon{qwen}Qwen 3.8 27B        & 3.3  & 0.0  & 27.1 & 13.9 & 64  & 23 & 72 & 9  & $+0.074$ (16) \\
\vendoricon{gemini}Gemini 3.8 Flash    & 0.0  & 0.0  & 6.4  & 0.0  & 9   & 33 & 56 & 11 & $-0.095$ (7)  \\
\vendoricon{gpt}GPT-6 Astra         & 23.0 & 21.1 & 53.5 & 25.0 & 158 & 15 & 84 & 5  & $-0.021$ (26) \\
\vendoricon{deepseek}DeepSeek V4.1 Flash & 2.0  & 2.6  & 31.2 & 16.7 & 65  & 42 & 69 & 15 & $-0.240$ (17) \\
\vendoricon{claude}Claude Opus 5.5     & 11.8 & 7.9  & 49.3 & 27.8 & 122 & 27 & 81 & 13 & $+0.018$ (27) \\
\midrule
\textbf{Pooled}     & --   & --   & --   & --   & 609 & 24 & 78 & 9  & -- \\
\bottomrule
\end{tabular}}

\end{table*}

\subsection{Interventions}
\label{app:res-ablation}
Both interventions use GPT-6 Astra and the opening requests of the main experiment.

\paragraph{Profile-visible intervention.}
\begin{table*}[p]
\centering
\caption{Effect of showing the agent the user's profile (GPT-6 Astra, 296 cells with identical opening requests). Each change is the profile-visible value minus the main-experiment value on the same cell. Satisfaction is the user's rating at delivery. Better / worse counts the action cells whose outcome changes. Bold: the interval excludes zero.}
\label{tab:ablation}
\resizebox{\textwidth}{!}{%
\begin{tabular}{@{}l c c c c c@{}}
\toprule
\multicolumn{1}{c}{\multirow[c]{2}{*}[-0.6ex]{\textbf{Group}}} & \multicolumn{3}{c}{\textbf{Information tasks}} & \multicolumn{2}{c}{\textbf{Action tasks}} \\
\cmidrule(lr){2-4}\cmidrule(lr){5-6}
 & $\Delta$ completion [95\% CI] & $\Delta$ satisfaction [95\% CI] & Median length (chars) & $\Delta$ completion [95\% CI] & Better / worse \\
\midrule
\textsc{expert}    & $-0.028$ [$-0.065$, $+0.009$] & $\mathbf{+0.34}$ [$+0.16$, $+0.53$] & 4148 $\to$ 4479 & $+0.083$ [$-0.028$, $+0.194$] & 4 / 1 \\
\textsc{senior}    & $\mathbf{-0.070}$ [$-0.115$, $-0.025$] & $-0.18$ [$-0.42$, $+0.05$] & 2688 $\to$ 2378 & $0.000$ & 0 / 0 \\
\textsc{young}     & $-0.015$ [$-0.068$, $+0.038$] & $-0.05$ [$-0.24$, $+0.13$] & 2750 $\to$ 2506 & $+0.083$ [$0.000$, $+0.194$] & 3 / 0 \\
\textsc{practical} & $+0.004$ [$-0.052$, $+0.065$] & $-0.13$ [$-0.32$, $+0.05$] & 3952 $\to$ 3520 & $-0.056$ [$-0.139$, $0.000$] & 0 / 2 \\
\midrule
\textbf{All}       & $\mathbf{-0.027}$ [$-0.051$, $-0.001$] & $-0.01$ [$-0.09$, $+0.09$] & 3469 $\to$ 3185 & $+0.028$ [$-0.007$, $+0.069$] & 7 / 3 \\
\bottomrule
\end{tabular}}

\end{table*}

Table~\ref{tab:ablation} compares the 296 persona cells with and without the profile. On information tasks the agent adjusts the level of detail. \textsc{expert} users receive longer answers, with median length increasing from 4,148 to 4,479 characters. Their satisfaction increases by $0.34$ (95\% CI $[+0.16,+0.53]$), while task completion changes by $-0.028$ (95\% CI $[-0.065,+0.009]$). \textsc{senior} users receive shorter answers, with the median falling from 2,688 to 2,378 characters. Their task completion falls by $0.070$ with CI $[-0.115,-0.025]$, as the number of satisfied requirements falls from 2.61 to 2.26, and their satisfaction does not rise, with a change of $-0.18$ and CI $[-0.42,+0.05]$. On action tasks only 10 of 144 cells change outcome, and requests that only ask for help reach a task completion of 0.17 with the profile against 0.15 without it. Overall, task completion changes by $-0.027$ on information tasks (95\% CI $[-0.051,-0.001]$) and $+0.028$ on action tasks (95\% CI $[-0.007,+0.069]$).

\paragraph{Forced-clarification intervention.}
\begin{table*}[p]
\centering
\caption{Effect of forcing the agent to ask at least one clarification question (GPT-6 Astra, 370 cells with identical opening requests). $\Delta$ is the forced value minus the main-experiment value on the same cell. Better / worse counts the cells whose score rises or falls, Asked is the share of cells with at least one question, and Steps is the mean number of agent steps.}
\label{tab:forced}
\resizebox{\textwidth}{!}{%
\begin{tabular}{@{}l r c c c c c@{}}
\toprule
\textbf{Slice} & \textbf{Cells} & \textbf{Task completion: main $\to$ forced} & \textbf{$\Delta$ [95\% CI]} & \textbf{Better / worse} & \textbf{Asked (\%)} & \textbf{Steps} \\
\midrule
Information, original & 38  & 0.762 $\to$ 0.776 & $+0.013$ [$-0.030$, $+0.054$] & 8 / 5   & 21 $\to$ 100 & 41.1 $\to$ 49.1 \\
Information, persona & 152 & 0.582 $\to$ 0.549 & $-0.032$ [$-0.065$, $+0.001$] & 28 / 42 & 23 $\to$ 100 & 37.0 $\to$ 50.0 \\
Action, original      & 36  & 0.583 $\to$ 0.472 & $-0.111$ [$-0.250$, $0.000$]  & 1 / 5   & 25 $\to$ 100 & 17.0 $\to$ 25.7 \\
Action, persona      & 144 & 0.278 $\to$ 0.285 & $+0.007$ [$-0.042$, $+0.056$] & 8 / 7   & 53 $\to$ 100 & 18.8 $\to$ 24.5 \\
\midrule
\textbf{All persona} & 296 & 0.434 $\to$ 0.421 & $-0.013$ [$-0.042$, $+0.018$] & 36 / 49 & 38 $\to$ 100 & 28.1 $\to$ 37.6 \\
\bottomrule
\end{tabular}}

\end{table*}

Table~\ref{tab:forced} compares all 370 cells of the main experiment with a rerun in which GPT-6 Astra must ask the user at least one question before it acts. Every persona cell now contains a question, against 38\% in the main experiment. The agent asks 2.4 questions and takes 9.5 more steps per cell. Yet task completion on persona cells does not recover, moving from 0.434 to 0.421 with $\Delta=-0.013$ and CI $[-0.042,+0.018]$. On information tasks the coverage of the requests does not rise either, moving from 0.757 to 0.740.

The required question even hurts the original requests. On action tasks, where the original instruction is already complete, task completion on original requests falls from 0.583 to 0.472, with $\Delta=-0.111$ and CI $[-0.250,0.000]$. Five cells turn from pass to fail and one from fail to pass. In these five, the persona agent answers with a constraint absent from the instruction, such as a year that differs from the site's data or a request to confirm by message before hiring. The agent follows it, and nothing is submitted. The gap between persona and original requests therefore narrows from $-0.241$ to $-0.207$ only because the original requests get worse. Requiring a question thus amplifies the persona agent's tendency to invent details.

\section{Case Studies}
\label{app:cases}
\begin{figure*}[t]
    \centering
    \includegraphics[width=0.97\linewidth]{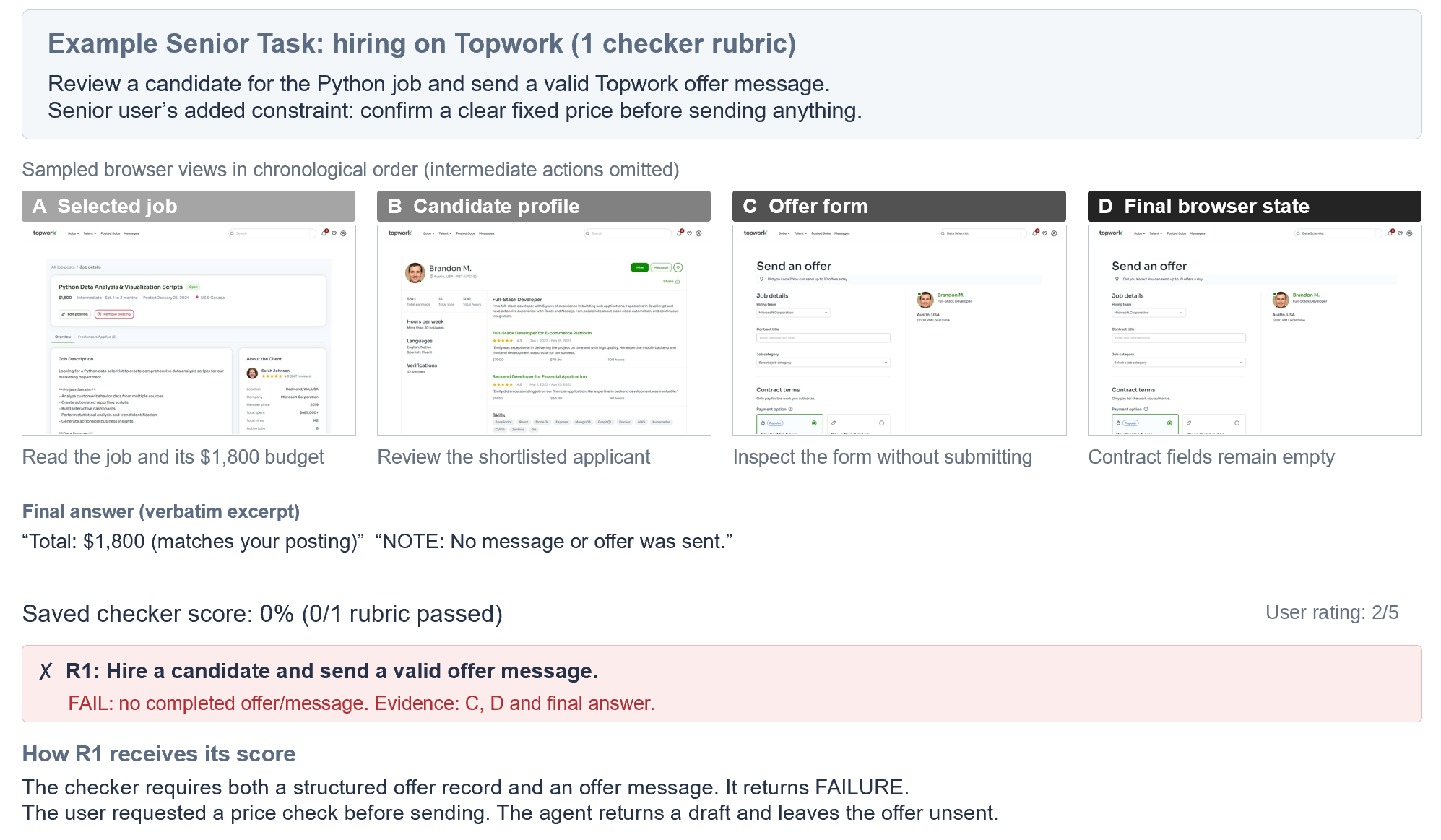}
    \caption{Senior persona case in PersonaCUABench, run by DeepSeek V4.1 Flash.  A cost-cautious user requests a fixed price before the agent sends an offer. The sampled views show job review, candidate review and an unsubmitted offer form. The agent returns a \$1,800 draft and states that nothing was sent. The original checker awards 0/1 for the unfinished offer, although the pause follows the boundary the user added. The user, who still wanted the hire, rates it 2/5. Note that, Figure~\ref{fig:overview} is run by MiniMax, with the same \textsc{senior} user which also stops before sending the offer and is rated 5/5.}
    \label{fig:case_study_senior}
\end{figure*}

\begin{figure*}[p]
    \centering
    \includegraphics[width=0.97\linewidth]{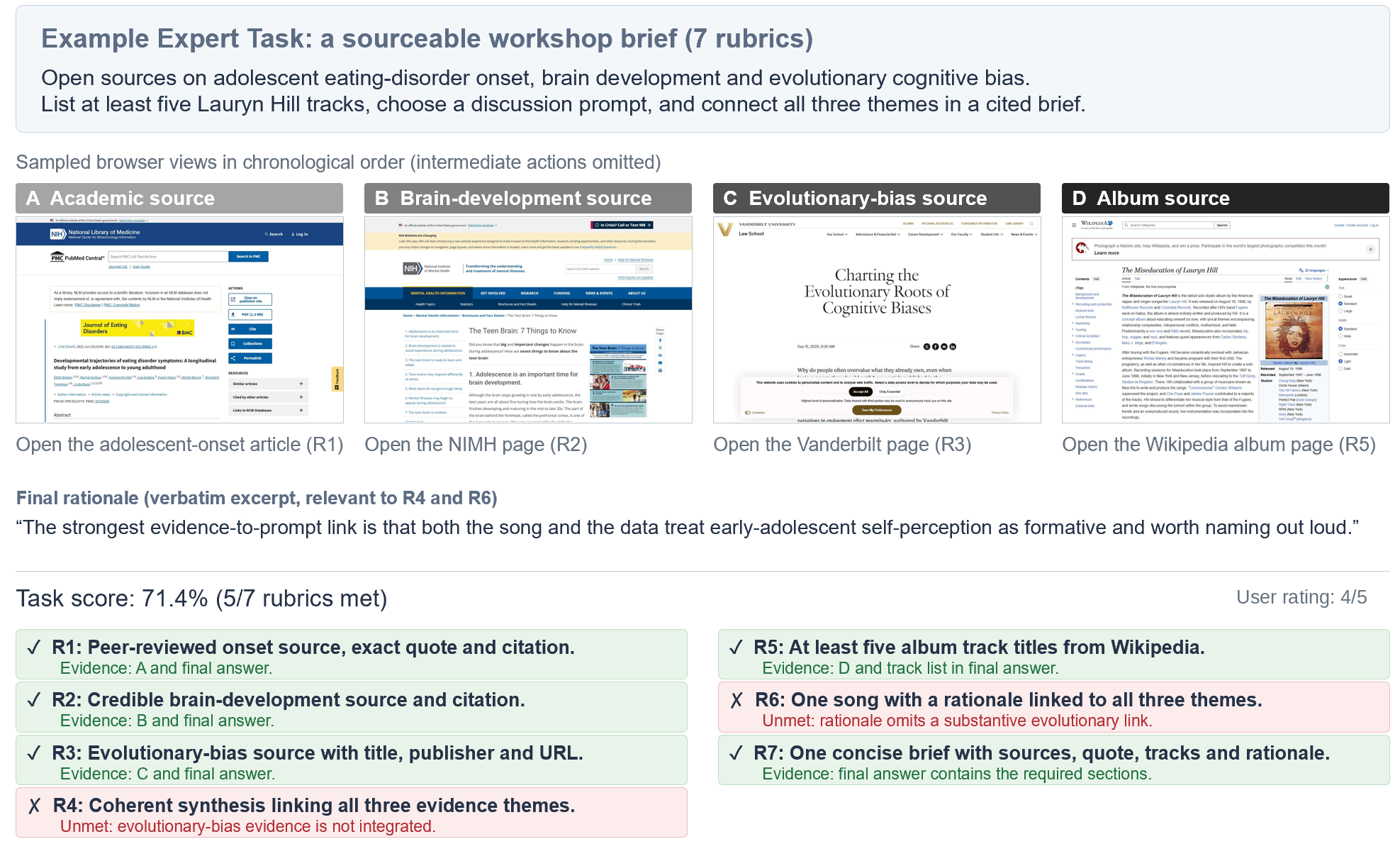}
    \caption{Expert persona case in PersonaCUABench run by DeepSeek V4.1 Flash. The user requests reusable sources and a cited workshop brief. The agent opens the three evidence pages and the album page, then supplies the sources and track titles. The saved rubric judge awards 5/7. R4 and R6 fail because the final synthesis and song rationale do not meaningfully integrate the evolutionary-bias evidence, despite opening its source.}
    \label{fig:case_study_expert}
\end{figure*}

\begin{figure*}[tp]
    \centering
    \includegraphics[width=0.97\linewidth]{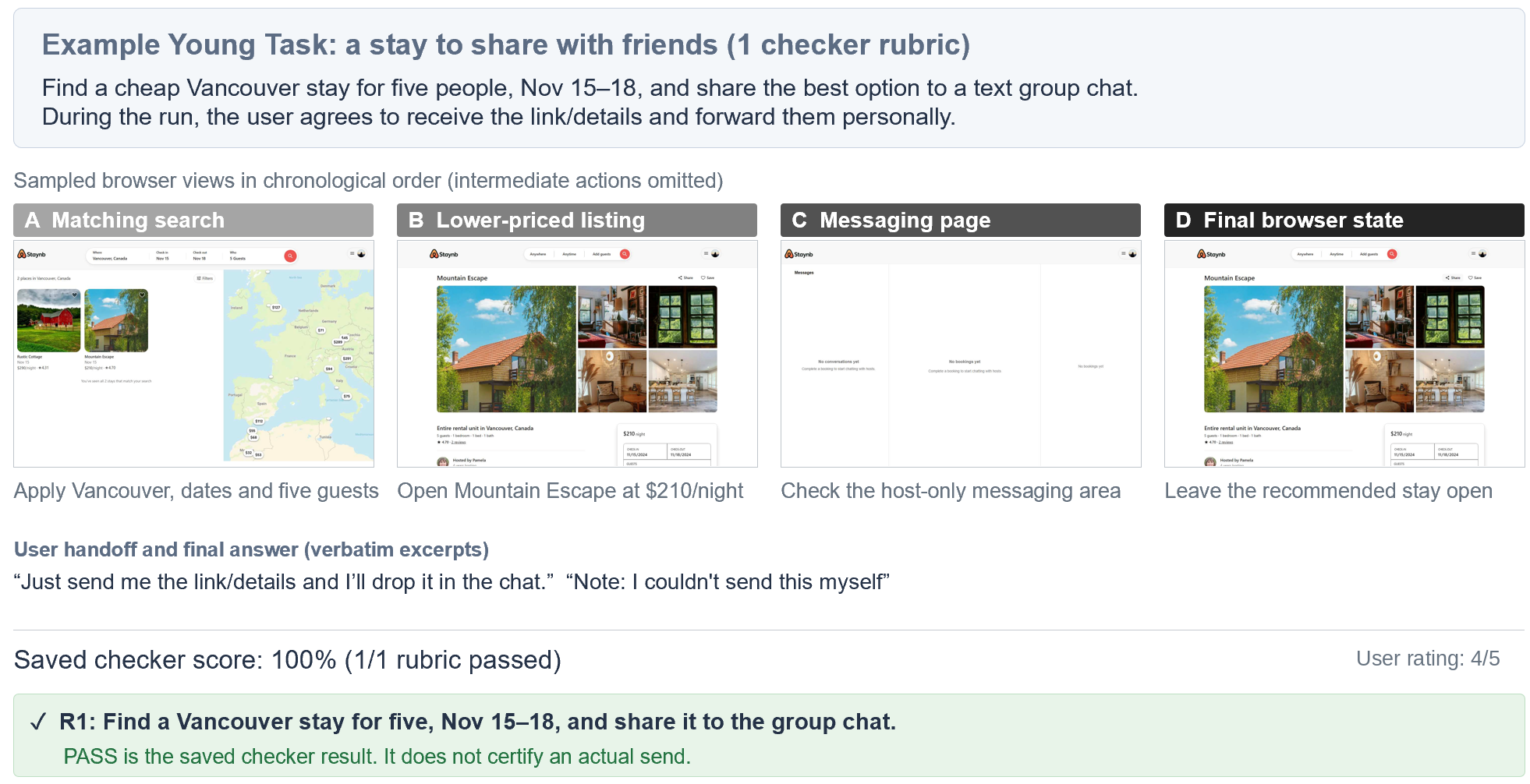}
    \caption{Young persona case in PersonaCUABench run by DeepSeek V4.1 Flash. The agent searches for a low-cost group stay, opens the cheaper listing and explores messaging. The user agrees to forward the link, and the final answer states that the agent did not send it. The saved checker awards 1/1, but its implementation checks search criteria only. This case exposes a gap between the checker’s pass and the original task’s group-chat delivery requirement.}
    \label{fig:case_study_young}
\end{figure*}

Figure~\ref{fig:case} in \S\ref{sec:res-case} shows the case of a \textsc{practical} user. Figures~\ref{fig:case_study_senior} to~\ref{fig:case_study_young} follow one cell for each of the other three groups, two on action tasks for \textsc{senior} and \textsc{young} users and one on an information task for an \textsc{expert} user. Each figure shows the task and the user's request, sampled browser views of the agent's main steps, an excerpt of its final answer, and how the cell is scored. Together with Figure~\ref{fig:case}, they show that the score and what the user wanted can diverge in both directions. A checker can fail an agent that stopped where the user asked it to (Figure~\ref{fig:case_study_senior}) or pass one that left part of the original task undone (Figure~\ref{fig:case_study_young}), and on information tasks the rubric judge gives partial credit when the answer falls short of the synthesis or decision that was asked for (Figures~\ref{fig:case_study_expert} and~\ref{fig:case}).

\end{document}